\documentclass[apj]{emulateapj}

\usepackage{amsmath}
\usepackage{epstopdf}
\usepackage{ulem}

\usepackage[colorlinks=true,citecolor=blue,filecolor=black,runcolor=blue,linkcolor=blue,urlcolor=blue]{hyperref}

\shorttitle{Milky-Way-Like and Massive-Galaxy Progenitors at $0.5<z<3.0$}
\shortauthors{MORISHITA et al.}

\begin{document}
\title{From Diversity to Dichotomy, and Quenching: Milky-Way-Like and Massive-Galaxy Progenitors at $0.5<z<3.0$}

\author{Takahiro~Morishita\altaffilmark{1,2}, 
	Takashi~Ichikawa\altaffilmark{1}, 
	Masafumi~Noguchi\altaffilmark{1}, 
	Masayuki~Akiyama\altaffilmark{1},
	Shannon~G.~Patel\altaffilmark{3},
	Masaru~Kajisawa\altaffilmark{4,5},
	and Tomokazu~Obata\altaffilmark{1}
	}	
\email{mtakahiro@astr.tohoku.ac.jp}
\altaffiltext{1}{Astronomical Institute, Tohoku University, Aramaki, Aoba, Sendai 980-8578, Japan}
\altaffiltext{2}{Institute for International Advanced Research and Education, Tohoku University, Aramaki, Aoba, Sendai 980-8578, Japan}
\altaffiltext{3}{Carnegie Observatories, 813 Santa Barbara Street, Pasadena, CA 91101, USA}
\altaffiltext{4}{Graduate School of Science and Engineering, Ehime University, Bunkyo-cho, Matsuyama 790-8577, Japan}
\altaffiltext{5}{Research Center for Space and Cosmic Evolution, Ehime University, Bunkyo-cho, Matsuyama 790-8577, Japan}

\begin{abstract}
\noindent
Using the $Hubble\ Space\ Telescope\ (HST)$/WFC3 and Advanced Camera for Surveys multi-band imaging data taken in CANDELS and 3D-HST, we study the general properties and diversity of the progenitors of the Milky Way (MWs) and local massive galaxy (MGs) at $0.5<z<3.0$, based on a constant cumulative number density analysis.
After careful data reduction and stacking analysis, we conduct a radially resolved pixel spectral energy distribution fitting to obtain the radial distributions of the stellar mass and rest-frame colors.
The stellar mass of MWs increases in self-similar way, irrespective of the radial distance, while that of MGs grows in an inside-out way where they obtain $\sim75\%$ of the total mass at outer ($>2.5$~kpc) radius since $z\sim2$.
Although the radial mass profiles evolve in distinct ways, the formation and quenching of the central dense region (or bulge) ahead of the outer disk formation are found to be common for both systems.
The sudden reddening of the bulge at $z\sim1.6$ and $z\sim2.4$ for MWs and MGs, respectively, suggests the formation of the bulge and would give a clue to the different gas accretion histories and quenching.
A new approach to evaluate the morphological diversity is conducted by using the average surface density profile and its dispersion.
The variety of the radial mass profiles for MGs peaks at higher redshift ($z>2.8$) and then rapidly converges to a more uniform shape at $z<1.5$, while that for MWs remains in the outer region over the redshift.
Compared with the observed star-formation rates and color profiles, the evolution of variety is consistently explained by the star-formation activities.

\end{abstract}

\slugcomment{Accepted for publication in The Astrophysical Journal}

\keywords{galaxies: high-redshift --- galaxies: evolution --- 
	galaxies: formation --- galaxies: general ---
	galaxies: structure --- Galaxy: evolution}


\section{INTRODUCTION}

The nature of high redshift (high-$z$) galaxies has been enthusiastically investigated for a decade.
Thanks to the recent progress of observation studies with the $Hubble\ Space\ Telescope$ ($HST$) and 10m-class ground-based telescopes, we have witnessed the clear and detailed structures of galaxies in the early Universe.
Galaxies in such an epoch are known to have quite different properties from local ones.
One example is the so-called $``red\ nugget,"$ which is passively evolving massive ($\sim10^{11} M_\sun$) galaxy with small radius $\sim1$~kpc (Daddi et al.~\citeyear{daddi07}; Trujillo et al.~\citeyear{trujillo07}; van Dokkum et al.~\citeyear{vandokkum08}; Damjanov~\citeyear{damjanov09}; Nelson et al.~\citeyear{nelson14}).
After a rapid formation phase at $z>2$, they are considered to increase stellar masses and sizes at relatively moderate speeds by accretions of satellite galaxies at the outer parts to evolve into massive elliptical galaxies in the local Universe (Naab et al.~\citeyear{naab09}; Oser et al.~\citeyear{oser10}; Trujillo et al.~\citeyear{trujillo12}).

On the other hand, the formation and evolution histories of the bulge-disk structure of disk galaxies, including the Milky Way (MW), are still matter of much debate; some studies explain them by in-situ star-formation with constant and moderate gas accretion from halo to central region (e.g., Kormendy \& Kennicutt~\citeyear{kormendy04}; Dekel et al.~\citeyear{dekel09}), while others use violent physical mechanisms, such as bar instability and giant clump migration (e.g., Noguchi~\citeyear{noguchi98},~\citeyear{noguchi99}; Debattista et al.~\citeyear{debattista06}).
Although more elaborate studies, especially by means of galactic archaeology, have been conducted based on the chemical abundance and kinematics of resolved stars, the details of the formation and evolution history of the MW are still unknown (e.g., Lee et al.~\citeyear{lee11}; Bovy et al.~\citeyear{bovy12}; Bergemann et al.~\citeyear{bergemann14}).

Recently, high-$z$ progenitor studies have shed light on the early evolution of the local galaxies.
By selecting progenitors that become, for example, massive elliptical galaxies in the local Universe, and observing their snapshots over a certain redshift range, recent studies have investigated their evolution histories.
Although there are a number of novel matching schemes to pick up progenitor galaxies (Ichikawa et al.~\citeyear{ichikawa07}; Conroy \& Wechsler~\citeyear{conroy09}; Guo et al.~\citeyear{guo10}; Tojeiro \& Percival~\citeyear{tojeiro10}), an example is based on a constant cumulative number density, which ranks galaxies by physical quantities, such as stellar mass.
The application of the analysis with constant cumulative number density to the MW progenitors (MWs) at high-$z$ was conducted by van Dokkum et al.~(\citeyear{vandokkum13}) with $HST$ imaging data.
By stacking the $H$-band images they derived average stellar mass profiles and showed that MWs had evolved in a self-similar way, where the mass growth rates at the inner and outer radii were comparable (see also Patel et al.~\citeyear{patel13b}).
Comparison with those of local massive galaxies (MGs) (van Dokkum et al.~\citeyear{vandokkum10}; Patel et al.~\citeyear{patel13}), which evolve in inside-out way, is also worth being mentioned (see also recent work by Papovich et al.~\citeyear{papovich15}).
However, the physical interpretation for the self-similar evolution was not pursued in past studies, whereas the inside-out evolution scenario was enthusiastically discussed (e.g., Oser et al.~\citeyear{oser10}; see also discussion in this study).

In order to avoid a significant contribution from young stars, near-infrared light is favorably used to derive galaxy stellar mass profiles (Szomoru et al.~\citeyear{szomoru13}).
Although $H$-band images (or longer wavelength, Kauffmann \& Charlot~\citeyear{kauffmann98}) are often applied to obtain the stellar mass profiles, more accurate ones could be achieved when the spatial variations of mass-to-light ratio, $M/L$, are reconstructed.
Moreover, with light profiles we could not know what physical mechanism(s) works in the galaxy evolution, whereas the studies of the local Universe seek for more specific evidences corroborated by, for example, mapping age distributions and chemical abundance of resolved stars.
Multi-object integral field units (IFUs) are also expected to achieve such scientific goals for galaxy evolution studies.
However, such spectroscopic observations are still expensive, especially for high-$z$ moderate mass galaxies of $\sim10^9M_\odot$.
Probing the general properties of star-formation and accumulation of stellar mass for such less massive galaxies still remains quite difficult.

The only alternative is the spatially resolved, pixel-by-pixel spectral energy distribution (SED) modeling, which we utilize with some modifications.
The pixel-by-pixel SED fit is not frequently used, because, for example, there are a limited number of available filter bands and a lack of spatial resolution with high signal-to-noise ratio (S/N).
The first attempt of spatially resolved SED fitting for high-$z$ star-forming and clumpy galaxies was made by Wuyts et al.~(\citeyear{wuyts12}) with the $HST$/Advanced Camera for Surveys (ACS) and WFC3 data (see also Zibetti et al.~\citeyear{zibetti09} for local galaxies).
Including follow-ups (e.g., Guo et al.~\citeyear{guo12}), these studies showed that the star-forming region of high-$z$ galaxies was extended to outer radii, which is consistent with the recent observations with IFU (F\"orster~Schreiber et al.~\citeyear{forster-schreiber14}; Genzel et al.~\citeyear{genzel14}).
We note, however, that most of the targets in the studies were massive ($M_*>10^{10}M_\odot$) and actively star-forming galaxies at the moderate redshift range ($0.7<z<1.5$).
Furthermore, the galaxy images are binned non-uniformly to obtain high S/Ns.
Such a binning method tends to blur morphological details.

In the context, we make use of the constant cumulative number density method to select the progenitors of the MW and local massive galaxy (MWs and MGs, respectively).
With highly resolved and deep multi-wavelength (optical to near-infrared) data provided by $HST$, we investigate the general properties and the diversity of the progenitor galaxies, such as SEDs, and resolved stellar mass and color profiles over the redshift of $0.5<z<3$, or, from the cosmic high-noon to dusk (Madau et al.~\citeyear{madau96}; Heavens et al.~\citeyear{heavens04}; Hopkins et al.~\citeyear{hopkins06}).
Stacking the sample in each redshift bin, with correction for position angle and axis ratio, allows us to obtain the average light profiles of both populations with sufficient S/N for each pixel.
With the multi-band average profiles, we derive radially resolved SEDs, which give a clue about to the radial properties and evolution histories of stellar mass and rest-frame colors.
Another novelty of this study is focusing on the diversity of galaxy morphology.
By making full use of the derived average profile and individual profiles, we evaluate the morphological varieties of the progenitors at each cosmological epoch for the first time.
These varieties are then investigated in conjunction with stellar mass, star-formation rate (SFR) and rest-frame colors.

The paper is organized as follows.
In Section~\ref{sec:sec_2} we describe the data and catalog, as well as the selection method for the progenitors.
The details of the resolved SED fitting and relevant reduction methods are followed in Section~\ref{sec:sec_3}.
The results and discussion are given in Sections~\ref{sec:sec_4} and \ref{sec:sec_5}.
The paper concludes with a summary in Section~\ref{sec:sec_6}.
Throughout the paper, we assume $\Omega_m$ = 0.3, $\Omega_\mathrm{\Lambda}$ = 0.7 and $H_0$ = 70 kms$^{-1}$Mpc$^{-1}$ for cosmological parameters and Chabrier (\citeyear{chabrier03}) initial mass function (IMF) for SED models.
We use the AB magnitude system (Oke \& Gunn~\citeyear{oke83}; Fukugita et al.~\citeyear{fukugita96}).

\section{DATA}\label{sec:sec_2}

\subsection{Sample Selection from the 3D-HST Catalog}\label{sec:sec_21}
The Cosmic Assembly Near-infrared Deep Extragalactic Legacy Survey (CANDELS; Grogin et al.~\citeyear{grogin11}; Koekemoer et al.~\citeyear{koekemoer11}) serves the multi-band optical to near-infrared imaging data of 2-10 orbits per pointing with the $HST$ for five well-known fields (GOODS-north and south, AEGIS, COSMOS and UDS).
The total survey area reaches $\sim$900~arcmin$^2$.
These deep and highly resolved imaging data enable us to obtain sufficient S/Ns for each pixel, even for high-$z$ ($>2$) galaxies, while eight-filter bands over optical to near-infrared wavelengths promise reliable stellar population syntheses.
We make use of a publicly available galaxy catalog by 3D-HST team (v.4.1; Skelton et al.~\citeyear{skelton14}).
Although the full description of the catalog is omitted, a brief summary is given here.
The photometric redshift was obtained with the EAZY code (Brammer et al.~\citeyear{brammer08}), which showed good consistency with previously obtained spectroscopic redshifts, $\Delta z/(1 + z)=0.003\--0.005$, for galaxies with $H\leq23$ (Skelton et al.~\citeyear{skelton14}).
The stellar mass was obtained with the FAST code (Kriek et al.~\citeyear{kriek09}), assuming a Chabrier (\citeyear{chabrier03}) IMF.
The best-fit SED was obtained with the $HST$ imaging data, Spitzer/IRAC mid-infrared data, and ground-based imaging data at the wavelengths from ultraviolet $U$-band to near-infrared $K$-band.
The SFR was derived from the rest-frame UV flux of the best-fit SED template and mid-/far-infrared data by {\it Spitzer}/MIPS.
In addition, medium-band imaging data are available for some regions, which increases the reliability of SED fitting.

From the 3D-HST catalog, we select the sample based on the spectroscopic redshift if available and photometric redshift for others, and stellar mass ($M_*$).
Further criteria for reliable photometry are set with $star\_flag$=0 and $use\_phot=1$, which limit the sample to the galaxies with $H\leq25$~mag in F160W-band.
Summing up image counts for higher redshift galaxies (therefore, with lower S/N) is an arduous and precarious task because wrong sky subtractions could lead to significantly biased results, especially for galaxies at the edge of observation fields, at nearby bright stars, in satellite trails, and in $\sim50$ pixel circular dead pixels on the WFC3/IR detector (referred to as "Death Star" by the 3D-HST team).
We visually exclude those erroneous galaxies.

\begin{figure}
\figurenum{1}
\begin{center}
\includegraphics[width=8cm,bb=3 14 707 530]{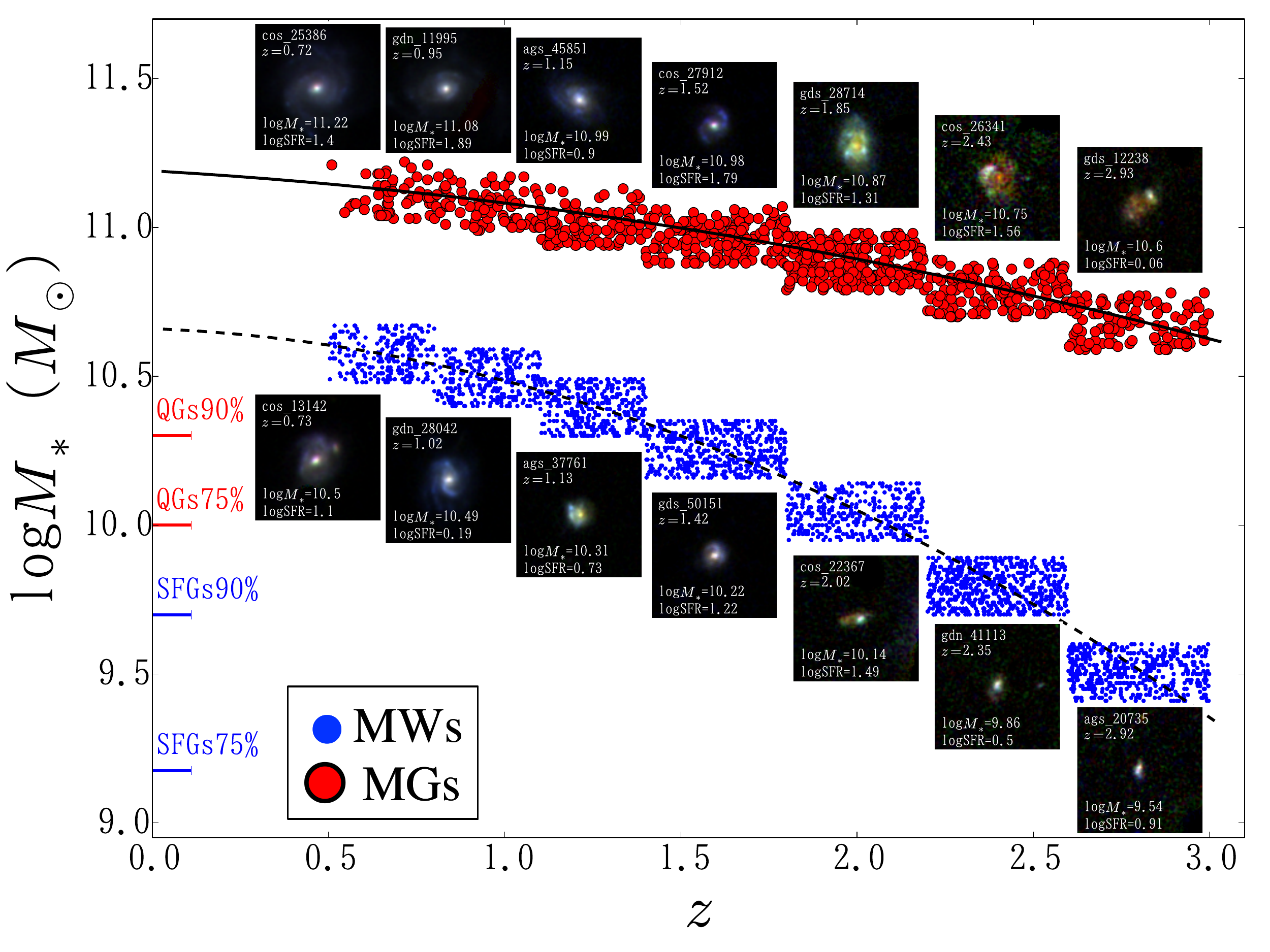}
\caption{
Progenitors of the Milky Way (MWs) and local massive galaxy (MGs) extracted by Eqs.~(\ref{eq_mw}) and (\ref{eq_mg})  (dashed and solid lines) shown with blue dots and red circles, respectively.
The limits of 90 and $75\%$ sample completeness for star-forming (SFGs, blue) and quiescent galaxies (QGs, red) are indicated on the ordinate.
A randomly selected sample for each population is depicted in each redshift bin, composited with F814W (blue), F125W (green) and F160W (red) of the $HST$ imaging data.
The stellar mass and star-formation rate (SFR) of each sample, which are taken from the 3D-HST catalog, are shown in the inset in unit of log$M_\odot$ and log$M_\odot$yr$^{-1}$.
All the samples are available at \href{http://www.astr.tohoku.ac.jp/~mtakahiro/mori15apj/mori_15_rgb.pdf}{http://www.astr.tohoku.ac.jp/$\sim$mtakahiro/mori15apj/mori\_15\_rg\\b.pdf}.
We follow MWs and MGs to $z\sim3$, where they had (only) $\sim7\%$ and $29\%$ of their present day masses ($\sim4.6\times10^{10}M_\odot$ and $1.5\times10^{11} M_\odot$), respectively.
}
\label{fig:fig_zm}
\end{center}
\end{figure}

The selected progenitor galaxies are shown in Fig.~\ref{fig:fig_zm}, with a randomly selected galaxy at each sample bin.
The sample is 90--95\% complete for galaxies with $M_* > 10^{10} M_\sun$ out to $z = 2.5$ (van der Wel et al.~\citeyear{vanderwel14}).
It is noted that the adopted criteria are sufficient to fully include the blue star-forming galaxies with $M_* \sim 10^9 M_\sun$ at $z\sim3$, while some of red quiescent counterparts might be missed.
However, since blue galaxies are dominant in numbers at higher redshift when the constant cumulative number density method is applied (van Dokkum et al.~\citeyear{vandokkum13}), the sample completeness is large enough for our purpose.
We show 90 and 75\% sample completeness limits for star-forming (SFGs, blue) and quiescent galaxies (QGs, red), which are derived in van der Wel et al.~(\citeyear{vanderwel14}), in the figure.

\subsection{Selection of the MW and MG Progenitors}\label{sec:sec_22}
We follow the constant cumulative number density criterion derived by van Dokkum et al.~(\citeyear{vandokkum13}) and Patel et al.~(\citeyear{patel13}) to select, at each redshift, galaxies that are supposed to become the MW-like and massive galaxies at $z\sim0$.
By using the constant number density for each redshift, van Dokkum et al.~(\citeyear{vandokkum13}) derived the stellar mass growth equation as a function of redshift for MWs ($n\sim1.1\times10^{-3}$~Mpc$^{-3}$),
\begin{equation}\label{eq_mw}
{\rm log} (M_*/M_\odot) = 10.66 - 0.045 z - 0.13 z^2,
\end{equation}
and Patel et al.~(\citeyear{patel13}) for MGs ($n\sim1.4\times10^{-4}$~Mpc$^{-3}$),
\begin{equation}\label{eq_mg}
{\rm log} (M_*/M_\odot) = 11.19 - 0.068 z - 0.04 z^2,
\end{equation}
where we convert the stellar mass with Kroupa~(\citeyear{kroupa01}) IMF ($M_\mathrm{*,K}$) into one with Chabrier IMF ($M_\mathrm{*,C}$) through log$(M_\mathrm{*,C}/M_\odot)~=~$log$(M_\mathrm{*,K}/M_\odot) - 0.04$ (Cimatti et al.~\citeyear{cimatti08}).
The equations above are both derived with the mass functions of Marchesini et al.~(\citeyear{marchesini09}) for high-$z$ galaxies, but at $z\sim0$ the former adopts the one of Moustakas et al.~(\citeyear{moustakas13}), while the latter of Cole et al.~(\citeyear{cole01}).
The difference between the two mass functions is negligible for the stellar mass range in the present study, and does not affect the sample selection.
According to the equations, expected stellar masses are log$(M_*/M_\odot)\sim$ 9.51, 9.79, 10.04, 10.25, 10.40, 10.49, and 10.57 for MWs and log$(M_*/M_\odot)\sim$ 10.68, 10.79, 10.89, 10.96, 11.00, 11.07, and 11.10 for MGs, respectively, at $\langle z\rangle\sim$ 2.8, 2.4, 2.0, 1.6, 1.2, 1.0, and 0.7.
\footnote{All the galaxies used for stacking are exhibited at \href{http://www.astr.tohoku.ac.jp/~mtakahiro/mori15apj/mori_15_rgb.pdf}{http://www.astr.tohoku.ac.jp/$\sim$mtakahiro/mori15apj/mori\_15\_rg\\b.pdf}.}
Although the redshift bins have different numbers of sample (all bins include $>80$ for MWs and $>10$ for MGs), the sample numbers are large enough to gain high S/N and for the following statistical discussion.
The numbers of galaxies used for the stacking are summarized in Table~\ref{tb:tb_stack}.
It should be noted that there is difference in the sample numbers of F125W and F160W, both of which are thought to have similar observation depth and field coverage.
By visually checking the imaging data, we find the difference originates from the satellite trails and bad pixels on the detector in either filter.
In addition, some galaxies are hidden under the sky background at shorter wavelengths, due to the ``$K$-correction," especially at higher redshift.
We exclude such galaxies from the final sample. 

Since galaxies could change their ranks for several reasons (e.g., major merging and quenching), we cut the sample with a range of 0.2~dex in the stellar mass (see Leja et al.~\citeyear{leja13}).
Although some of previous studies adopted different number densities for the progenitor selection (Leja et al.~\citeyear{leja13}; Ownsworth et al.~\citeyear{ownsworth14}), we verified that adopting such number densities do not change our results.
The sample in this study is selected with box of $z$ and $M_*$, as well as previous studies, rather than in curved box where stellar mass criteria evolve continuously with redshift.
Although the sample selection with box could bias the sample if the box size is large, in the present study we find no significant difference in the median values of the parameters (redshift, stellar mass, and SFR) between the two selection criteria.
The median values for $M_*$, $z$, and SFR are also summarized in Table~\ref{tb:tb_phys}.

\section{METHODOLOGY}\label{sec:sec_3}

\begin{figure*}
\figurenum{2}
\begin{center}
\includegraphics[width=12cm, bb=0 0 672 704]{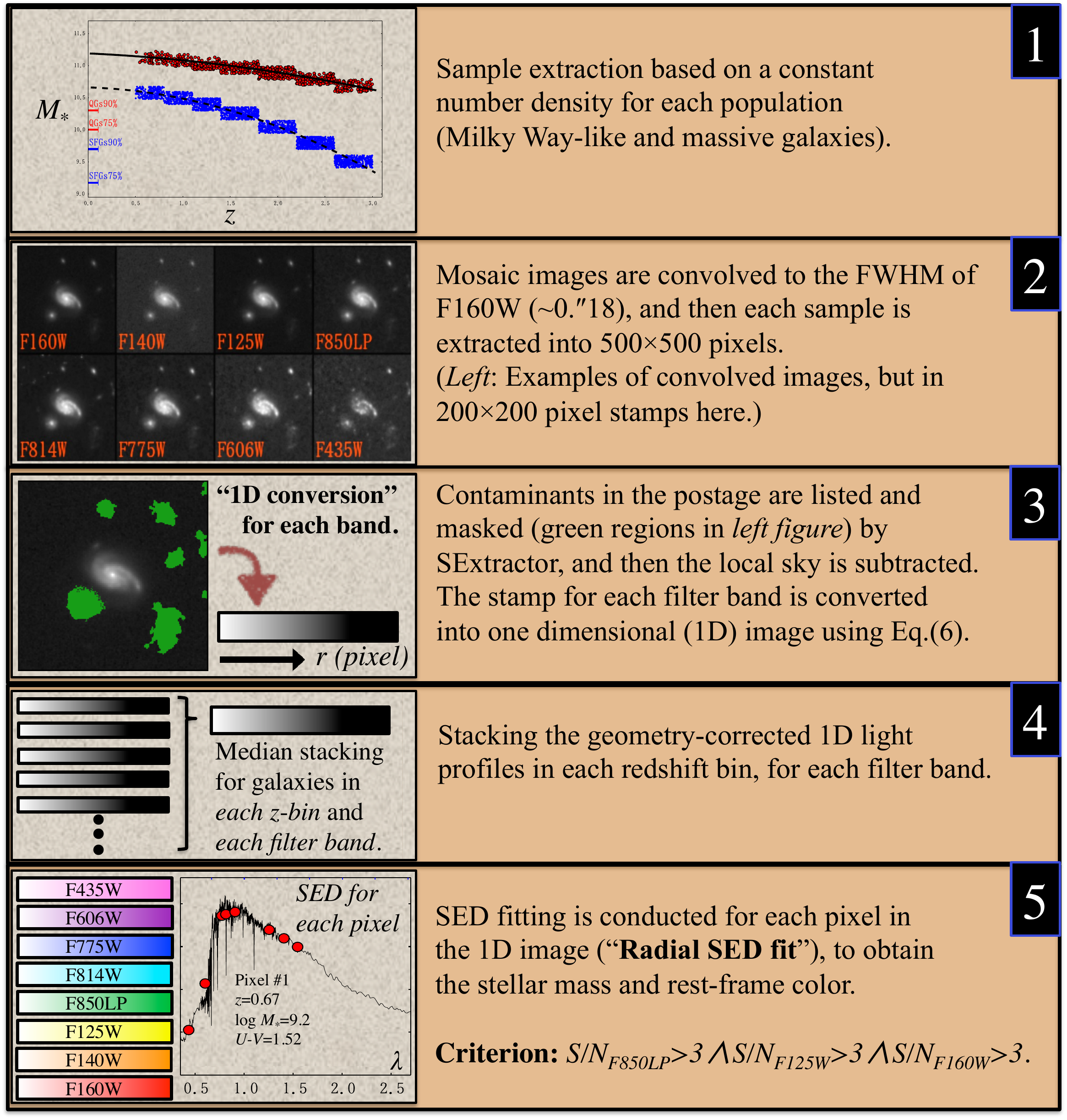} 
\caption{
Visual summary of the methodology in Section~\ref{sec:sec_3}: {\bf 1.}~sample selection (Section~\ref{sec:sec_22}), {\bf 2.}~PFS matching (Section~\ref{sec:sec_31}), {\bf 3.}~sky subtraction and 1-D conversion (Section~\ref{sec:sec_32}), {\bf 4.}~median stacking the 1-D profiles in each redshift bin for each filter band (Section~\ref{sec:sec_33}), and {\bf 5.}~Radial SED fit for each pixel (Sections~\ref{sec:sec_34} and \ref{sec:sec_42}).
We obtain the {\it radially resolved} stellar mass and rest-frame colors for MWs and MGs.
}
\label{fig_method}
\end{center}
\end{figure*}

Here we introduce the method of radially resolved SED fit, or radial SED, to derive the stellar mass and rest-frame colors for each pixel of the galaxy radial profiles.
We also visually summarize the contents in this Section in Fig.~\ref{fig_method}.

\subsection{Data Assessment}\label{sec:sec_31}
We use the imaging data of WFC3/IR (science, weight, and exposure images) reduced by the 3D-HST team (Skelton et al.~\citeyear{skelton14}).
For the data taken with ACS, however, we have no access to the exposure maps made by the 3D-HST team.
Therefore, for ACS imaging data we assume the exposure time for each, based on the scheduled observation time (A. M. Koekemoer 2014, private communication).

The error estimate of the flux is one of the most important processes for deriving SEDs.
Calculation for the $i$th pixel error, $E_i$, is made by the following equation,
\begin{equation}
E_i = \sqrt{{\sigma_{w,i}}^2+{\sigma_{o,i}}^2}.
\end{equation}
The first term of the right side, ${\sigma_{w,i}}^2$, is the $i$th pixel value of the variance map (or inverse weight map, $\_wht.fits$), in units of (e$^-$/s)$^2$, defined as
\begin{equation}\label{eq:eq_er}
\sigma_{w,i}^2= {{(D_i+B_i)+{N_i\sigma_\mathrm{read}}^2}\over{{t_i}^2}},
\end{equation}
where $D_i$ is the total accumulated dark current signals and $B_i$ is the total accumulated background level during the exposure on the $i$th pixel, both in unit of e$^-$.
$N_i$ is the total number of exposures, $\sigma_\mathrm{read}$ is the readout noise, and $t_i$ is the total exposure time.
We note that the variance map does not include the Poisson noise of the objects, but only background (sky, dark current, and readout) noise, including the pixel correlation noise during the drizzle task (see Casertano et al.~\citeyear{casertano00} for details).
Therefore, we need to add the Poisson noise for the object, ${\sigma_{o,i}}^2$ in unit of (e$^-$/s$^2$), as defined by
\begin{equation}
{\sigma_{o,i}}^2 = F_i/t_i^2,
\end{equation}
where $F_i$ denotes the total photon count (e$^-$) on the $i$th pixel of the reduced imaging data.

\begin{figure}
\figurenum{3a}
\begin{center}
\includegraphics[width=8cm,bb=0 0 172 98]{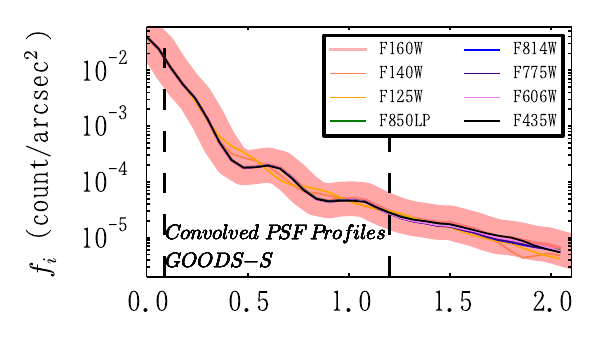}
\includegraphics[width=8cm,bb=0 0 172 98]{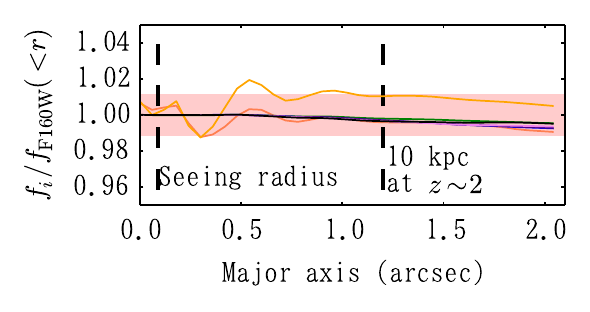}
\caption{
Examples of the PSF profiles (stacked stars provided by the 3D-HST team) convolved to match to that of the F160W image (red shade).
Vertical dashed lines represent the seeing radius ($\sim0.\!\arcsec09$) and physical scale ($\sim10$~kpc at $z\sim2$).
The width of the red shade in the bottom panel represents an error range within a $\pm1\%$ difference from the one of F160W.
}
\label{fig_psf}
\end{center}
\end{figure}

The science images are then convolved with point-spread function (PSF) to be blurred to the FWHM of F160W ($\sim0.\!\arcsec18$), while the error maps are convolved in quadrature.
The convolution is conducted with convolution kernels generated by the ``CLEAN" algorithm (H\"ogbom~\citeyear{hogbom74}), where PSF images provided by the 3D-HST team (median stacked stars) are used.
Examples of PSF profiles convolved to match the F160W PSF are shown in Fig.~\ref{fig_psf}, where we see that all the profiles are in good agreement, within $\sim2\%$ difference, which is much smaller than the photometric error (see also appendix).
At last we have two convolved image data, one for the science map (where the $i$th pixel has $F_i$) and the other for the error map ($E_i$), from which we calculate the S/N of the $i$th pixel as $F_i/E_i$, for each filter band.

\subsection{Conversion of Two-dimensional (2-D) Image to Onedimensional (1-D) Radial Light Profile}\label{sec:sec_32}
Prior to stacking the galaxies in redshift bins, we convert the 2-D imaging data into 1-D radial profiles (hereafter, 1D-conversion) to obtain the average radial profile.
We first extract the 2-D image of 500$\times$500~pixel, or $\sim$250$\times$250~kpc at $z\sim2$, from the original mosaic images to subtract the local sky background.
Since the calculated sky background calculated would sometimes be overestimated by bright neighboring objects (e.g., H\"{a}u{\ss}ler et al.~\citeyear{haussler07}), the following procedures are conducted.
First, we aggressively mask all objects detected with SExtractor (Bertin \& Arnouts~\citeyear{bertin96}) and the parameters in Table~\ref{tb:tb_sex}.
The masked pixels are not included in the calculation of sky.
We then estimate the local sky value from the median of the masked stamp.
Although wrong sky subtractions would give systematically biased results, especially for the ground-based near-infrared observations, the sky subtraction for the imaging data with $HST$ would not affect the results in this study because sky is negligibly fainter than the stacked profiles.
It is noted that applying the 1D-conversion after stacking 2-D images could yield false spatial asymmetricities.

\begin{figure}
\figurenum{3b}
\begin{center}
\epsscale{0.8}
\includegraphics[width=8cm,bb=0 0 576 288]{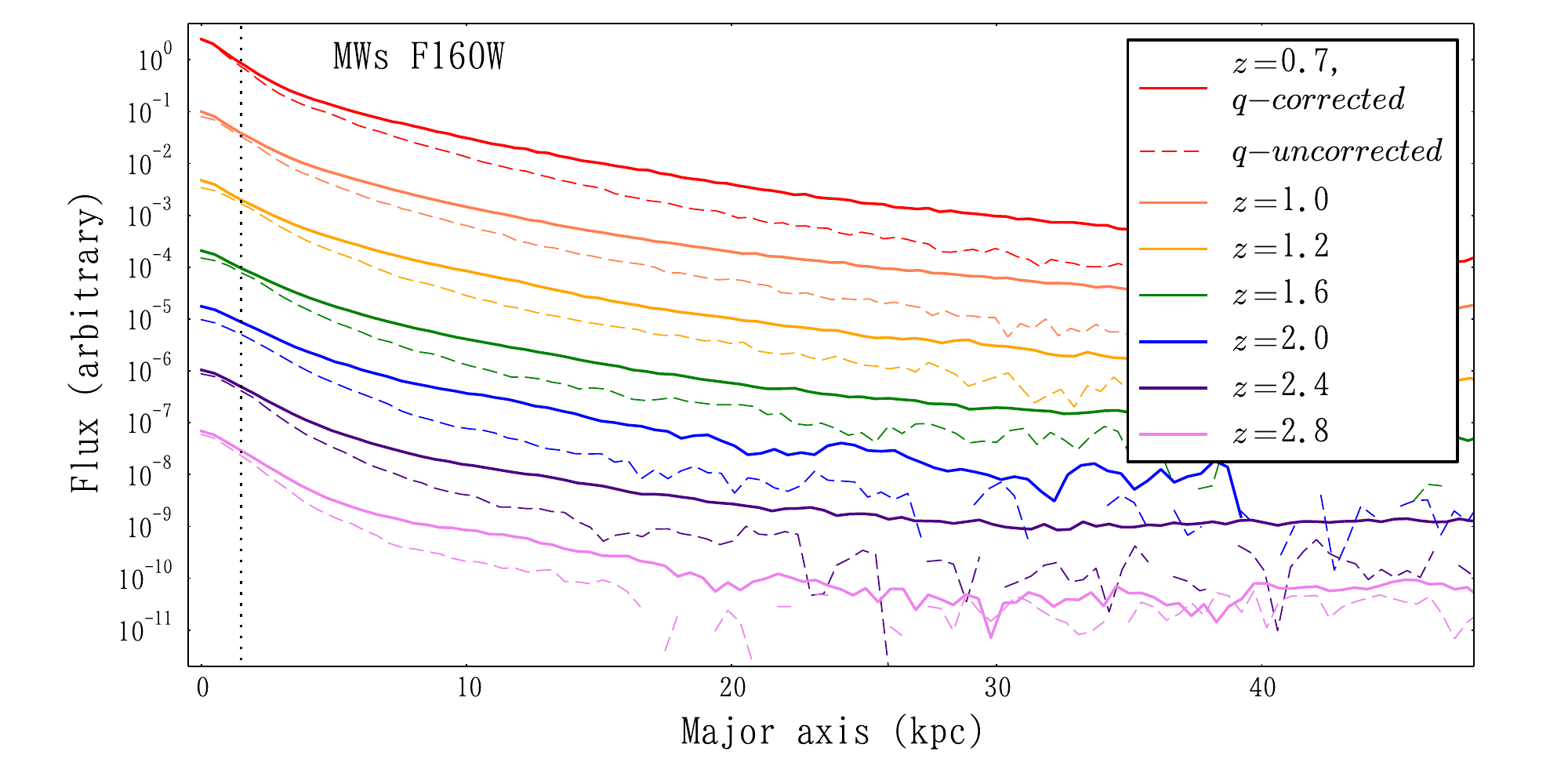}
\includegraphics[width=8cm,bb=0 0 576 288]{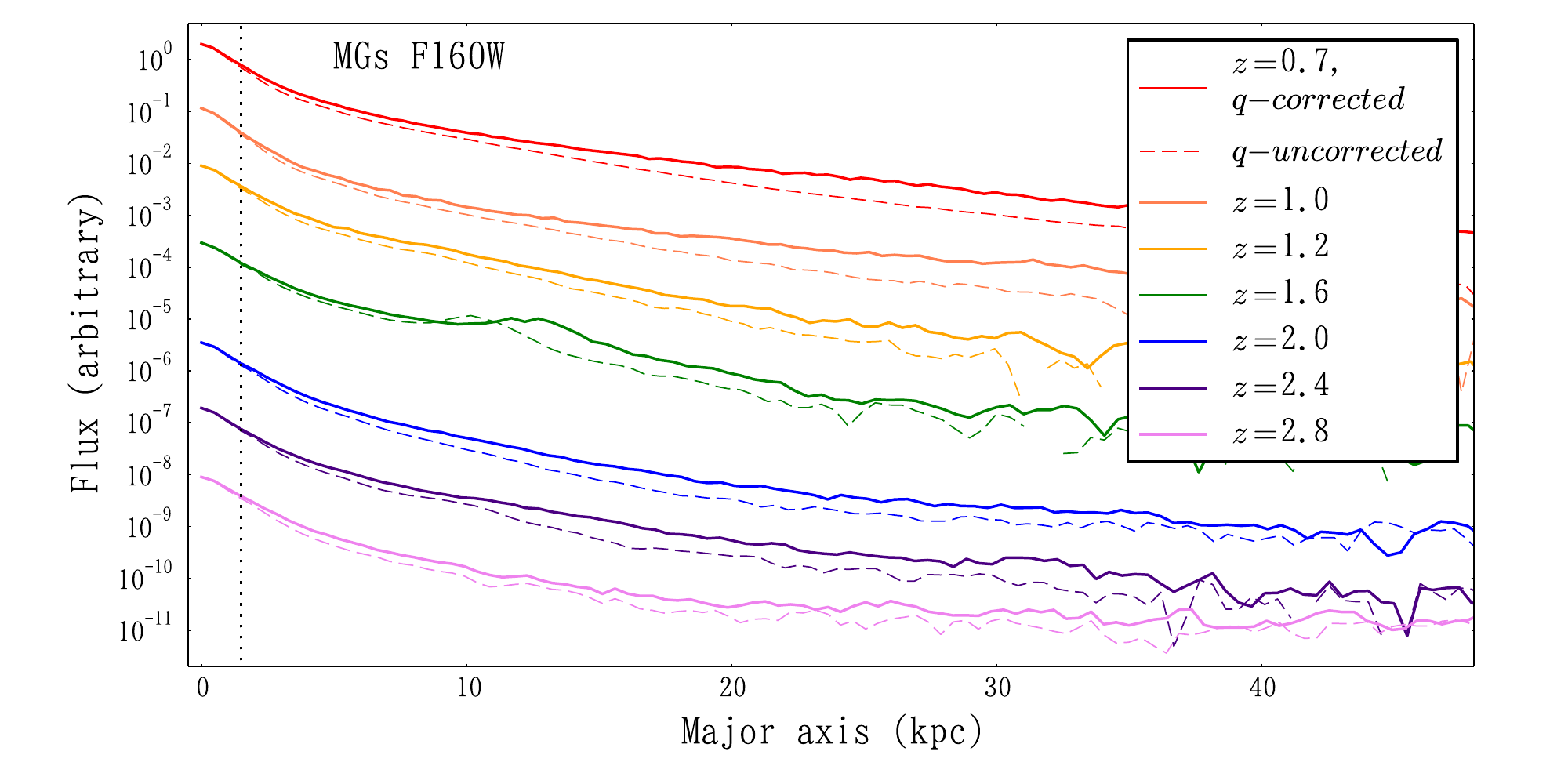}
\caption{
Comparison of the two median light (F160W) profiles that are stacked with (solid) and without inclination correction (dashed).
The profile flux is arbitrarily shifted for each redshift.
The profiles are similar in the central region ($<2$~kpc), but show deviation at the outer radius, for $\sim30\%$ at $\sim10$~kpc, which would bias the profile more centrally concentrated in case of uncorrected stacking.
Vertical dotted lines represent seeing radius.
}
\label{fig_inc}
\end{center}
\end{figure}

In the 1-D conversion, we correct the effect of galaxy inclination.
We obtain the projected axis ratio and apply the following equation (see also Bertin \& Arnouts~\citeyear{bertin96}) to convert 2-D imaging data into the inclination corrected 1-D profile,
\begin{equation}
\begin{split}
r^2(\theta, q) = (\cos^2\theta+\sin^2\theta/q^2)(x-x_0)^2 \\
+ (\sin^2\theta+\cos^2\theta/q^2)(y-y_0)^2 \\
+2\cos\theta(1-1/q^2)(x-x_0)(y-y_0),
\end{split}
\end{equation}
where $\theta$ and $q$ are position angle and axis ratio derived with SExtractor.
The correction has not been included in the previous studies, whereas stacking 2-D images without any geometrical corrections could, as is evident, result in morphologically biased (more centrally concentrated) images.
We show the comparison of the two median stacked profiles, which are inclination corrected and uncorrected, in Fig.~\ref{fig_inc}.
(For the details of the median stacking method, see the following subsection.)
We see larger discrepancy at the outer radius, where uncorrected profiles undergo for $\sim30\%$ of the corrected ones at any redshifts, while we see similar profiles at the inner radius ($<2$~kpc).
This is because the inclination effect is smaller at the rounder inner region.
The effects would worsen S/N and artificially bias the stacked profiles toward more centrally concentrated ones.
Accordingly, in what follows we make use of the 1-D profiles with the axis ratio and position angle corrections to derive the stacked average profiles.

\subsection{Stacking Analysis}\label{sec:sec_33}
In this subsection we stack the 1D light profiles in each redshift bin to obtain the composite radial profiles.
The procedure gives sufficient S/Ns for each pixel, even for the outer envelope of the high-$z$ galaxies.
Furthermore, the stacking analysis provides us with the radial profiles for all filter bands, because an observation field that lacks some filter bands (for example, COSMOS field lacks F435W, F775W, and F850LP) can be compensated by the data taken in other fields with those bands.
The full coverage of the observed bands would warrant reliable SED models.

The stacking of galaxies is conducted for each filter band as follows.
We use the inclination corrected 1-D imaging data, prepared in the previous section.
The background objects and contaminants are masked, and the masked pixels are not used for the median calculation when stacking.
Since the 1-D conversion relies on the center position in the original 2-D image, we use galaxy positions listed in the 3D-HST catalog (see third and force column in Fig.~\ref{fig_method}) for the stacking.

In addition to the local sky subtraction in the previous subsection, we compare the sky values of the final stacked profile by estimating at different radii ($50<r<75, 75<r<100, 100<r<125$ and $125<r<150$ pixel), and then subtracting the median values from the stacked profile, although they are negligible (see appendix for the detailed discussion).

Another concern of stacking analysis is the luminosity bias due to redshifts and stellar masses.
Since we divide the sample into finite redshift and stellar mass bins, the luminosities differ depending on redshifts and stellar masses, which would give rise to bias when stacking profiles.
To avoid the effect, we correct the luminosity for each galaxy by multiplying the following constant,
\begin{equation}\label{eq_k}
K(\langle z\rangle,\langle M_*\rangle)={(1+z_\mathrm{obs})^3\over{(1+\langle z\rangle)^3}}{\langle M_*\rangle \over{M_\mathrm{*,obs}}},
\end{equation}
where ``obs" represents the observed quantity and the brackets are the medians of each redshift bin. 
It should be noted that the correction includes the $K$-correction.
Since the SED fit (implicitly) includes the $K$-correction when applying the redshift shift, the latter part of Eq.~\ref{eq_k} ($\langle M_*\rangle/M_{*,\mathrm{obs}}$) corrects two effects: the $K$-correction and stellar mass difference.
The remaining part of the equation ($(1+z_\mathrm{obs})^3/(1+\langle z\rangle)^3$) corrects the luminosity distance.

\begin{figure}
\figurenum{4}
\begin{center}
\epsscale{0.8}
\includegraphics[width=9cm,bb=0 0 576 432]{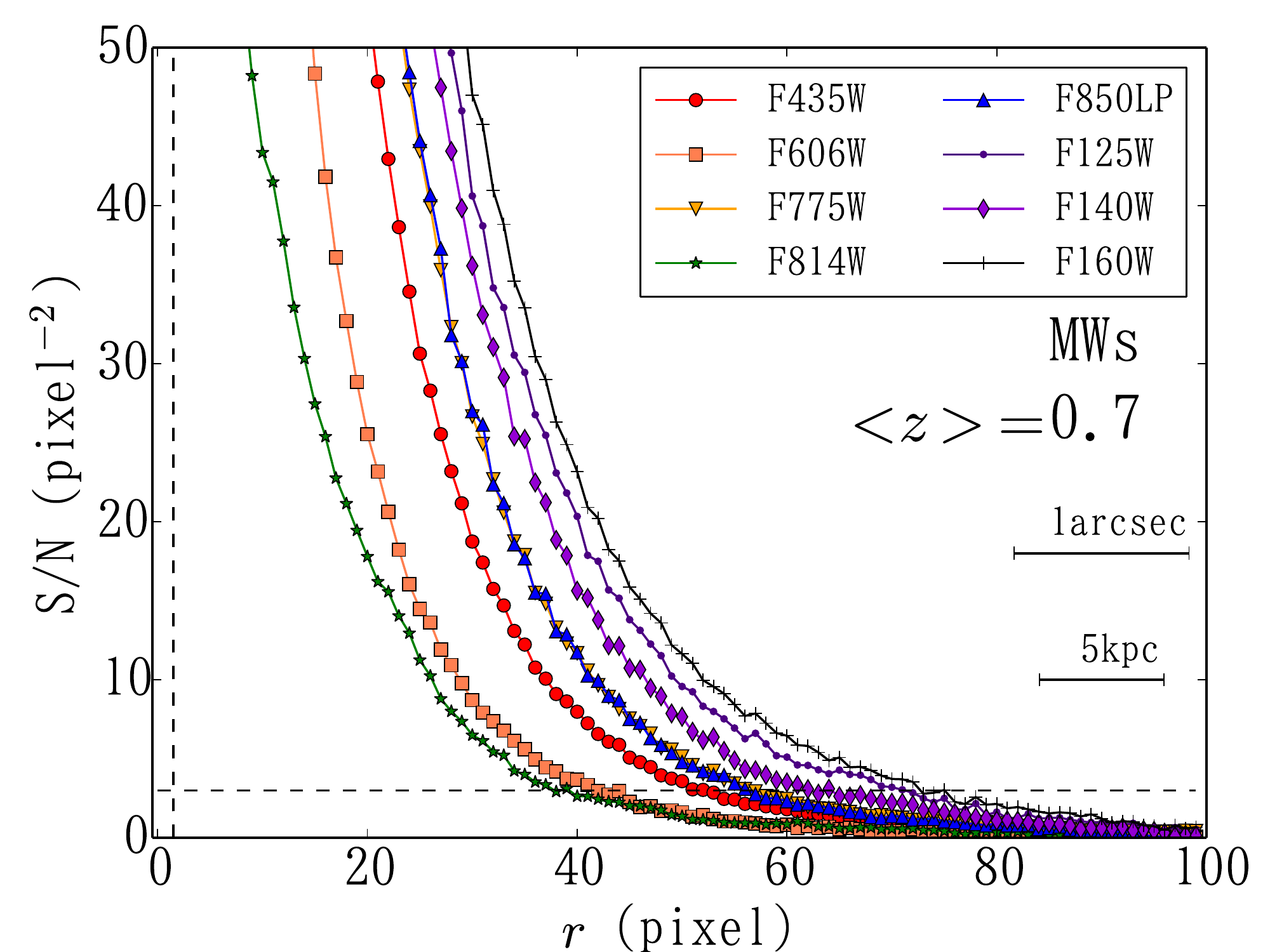}
\includegraphics[width=9cm,bb=0 0 576 432]{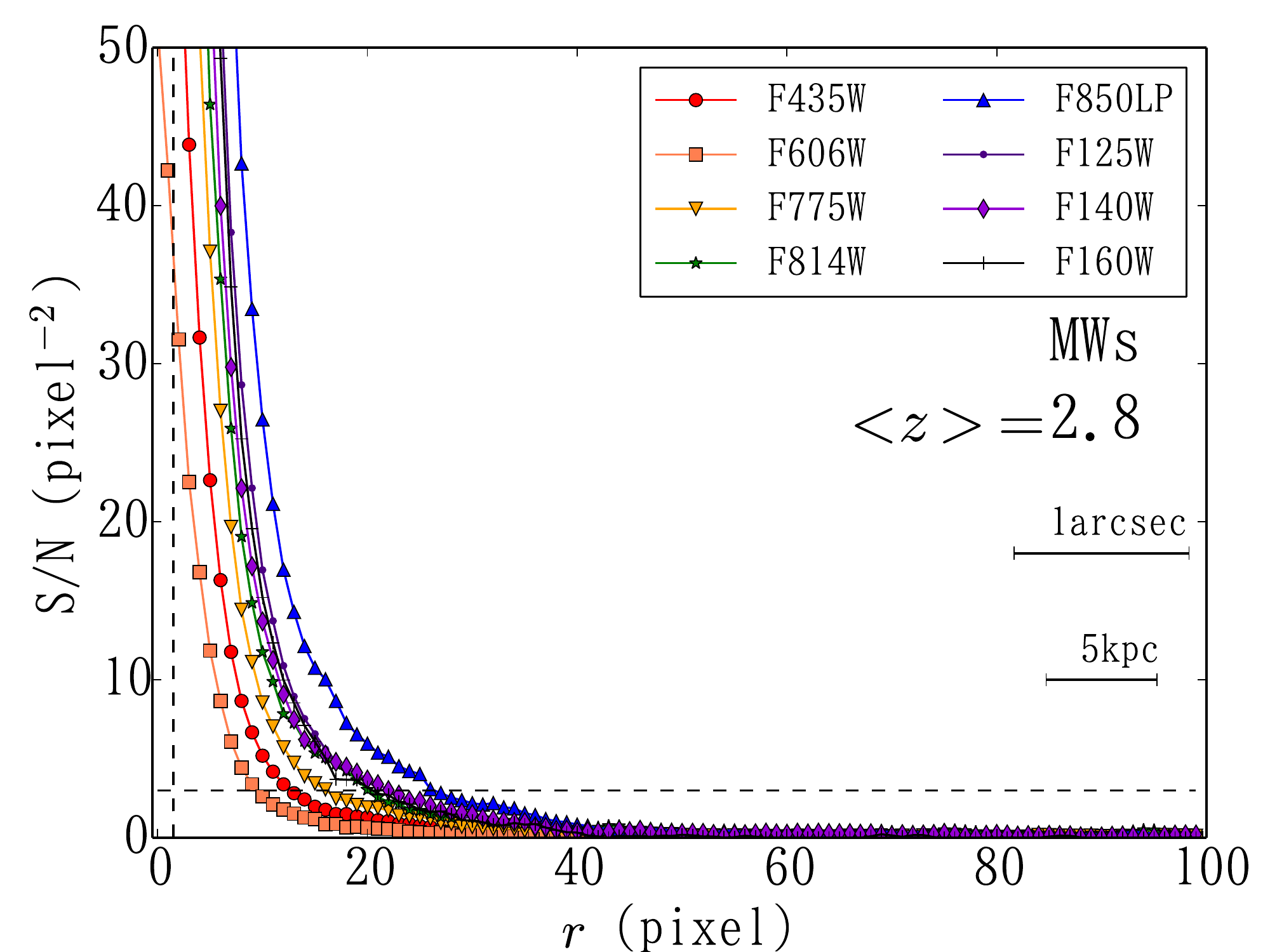}
\caption{
Examples of the radial signal-to-noise (S/N) profiles for the stacked MWs at $\langle z\rangle=0.7$ (top) and $2.8$ (bottom).
The vertical dashed line represents the PSF radius (FWHM/2) for each redshift.
The horizontal dashed lines represent the criterion, S/N~$>3$, for radial SED analysis.
We conduct the SED fit for each radial pixel with S/N~$>3$.
}
\label{fig:fig_sn}
\end{center}
\end{figure}

Figure~\ref{fig:fig_sn} shows examples of radial S/N profiles for the stacked galaxies at $z\sim0.7$ and $z\sim2.8$.
We find that the region with S/N ($>3$) reaches at least $\sim10$~pixel ($\sim5$~kpc at $z\sim2$) from the galaxy center, even in the highest redshift bin.
The number of stacked samples for each redshift bin and filter is summarized in Table~\ref{tb:tb_stack}.

\subsection{SED Fitting}\label{sec:sed}\label{sec:sec_34}
We calculate radially resolved SEDs for the stacked radial profiles using FAST (Kriek et al.~\citeyear{kriek09}) and the stellar population models of GALAXEV (Bruzual \& Charlot~\citeyear{bruzual03}), assuming solar metallicity and Chabrier (\citeyear{chabrier03}) IMF.
The MW dust attenuation of Cardelli et al.~(\citeyear{cardelli89}) is adopted in the range of $0\leq A_\mathrm{V}\leq4.0$~mag by a step of 0.1.
The age is set to range from 0.1 to 10~Gyr, or up to the age of the Universe at the galaxy redshift.
Redshifts are set to the median values of the sample bins.

The star-formation history is assumed to be an exponentially declining model, SFR~$\propto \exp(-t/\tau)$, where $t$ is the time since its star-formation starts and $\tau$ is the $e$-folding timescale of the SFR.
It is noted that the exponential model was first proposed for local elliptical galaxies, and recent studies have shown that the delayed and truncated models would be consistent with observations of high-$z$ star-forming galaxies (Maraston et al.~\citeyear{maraston10}; Barro et al.~\citeyear{barro13}).
However, in this study we adopt the exponential model because with only the photometric data we are not able to discriminate these star-formation histories.
Instead, we adopt a dispersion of the best-fit stellar masses derived with the three star-formation histories (exponential, delayed, and truncated models) as a typical error of the stellar mass of the exponential model.
The error ($\sim0.1$~dex) is much larger than those (e.g., due to the sky subtraction error).
See the appendix for the discussion of the uncertainties in the stellar mass.
It is noted that stellar mass is the most reliable parameter to be derived in SED fitting, as demonstrated by Wuyts et al.~(\citeyear{wuyts12}), whereas the other parameters wander through the complex parameter space (e.g., well-known age-metallicity-dust degeneracy; see also Papovich et al.~\citeyear{papovich01}; Shapley et al.~\citeyear{shapley05}).
The visual summary of the methodology described here is shown in Fig.~\ref{fig_method}.
In the next sections, we show the results of the radial SED fitting for MWs and MGs.

\section{RESULTS}\label{sec:sec_4}

\subsection{Integrated SEDs of Stacked Samples}\label{sec:sec_41}
\begin{figure*}
\figurenum{5}
\begin{center}
\includegraphics[width=\textwidth,bb=0 0 576 288]{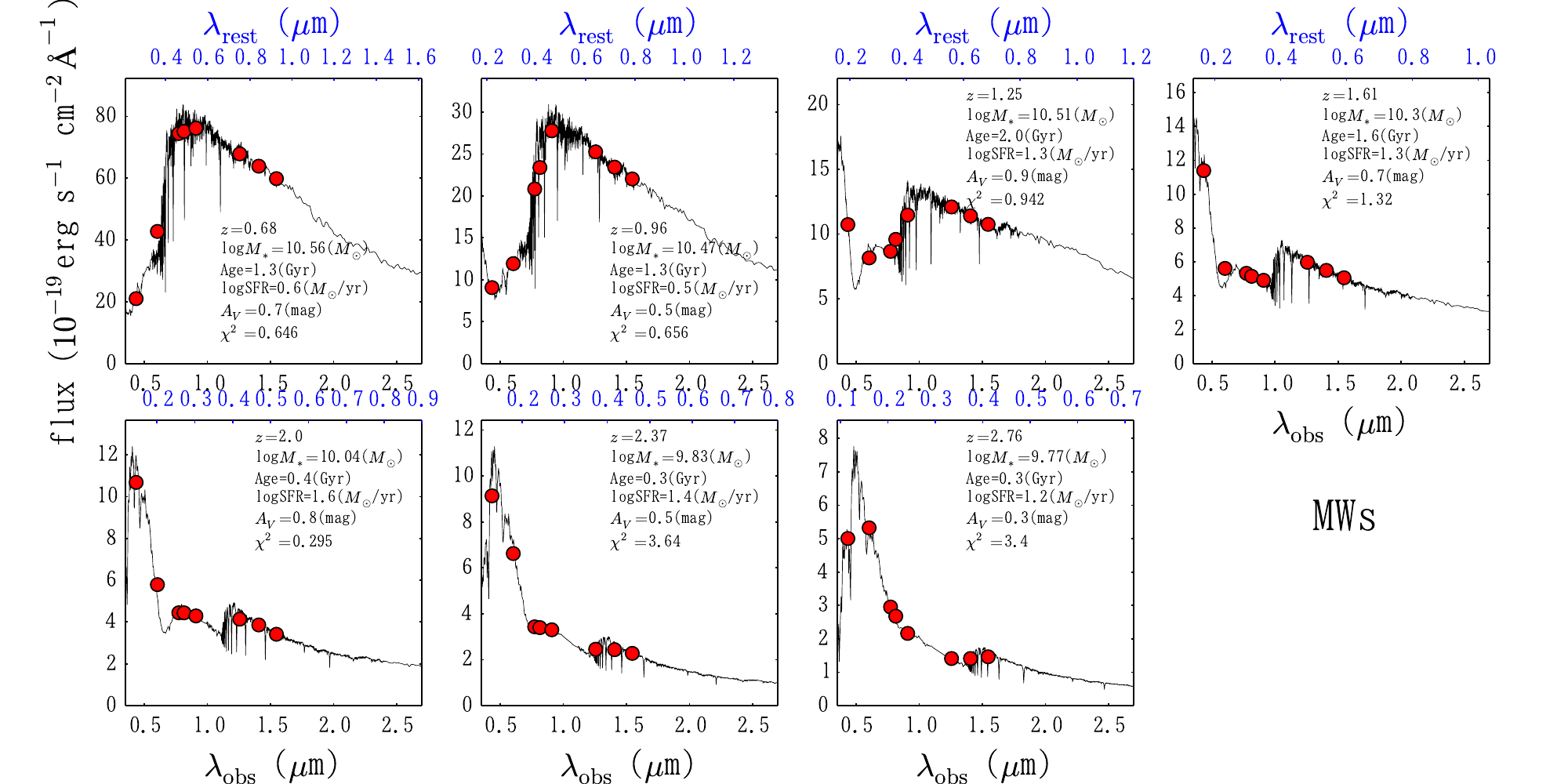}
\includegraphics[width=\textwidth,bb=0 0 576 288]{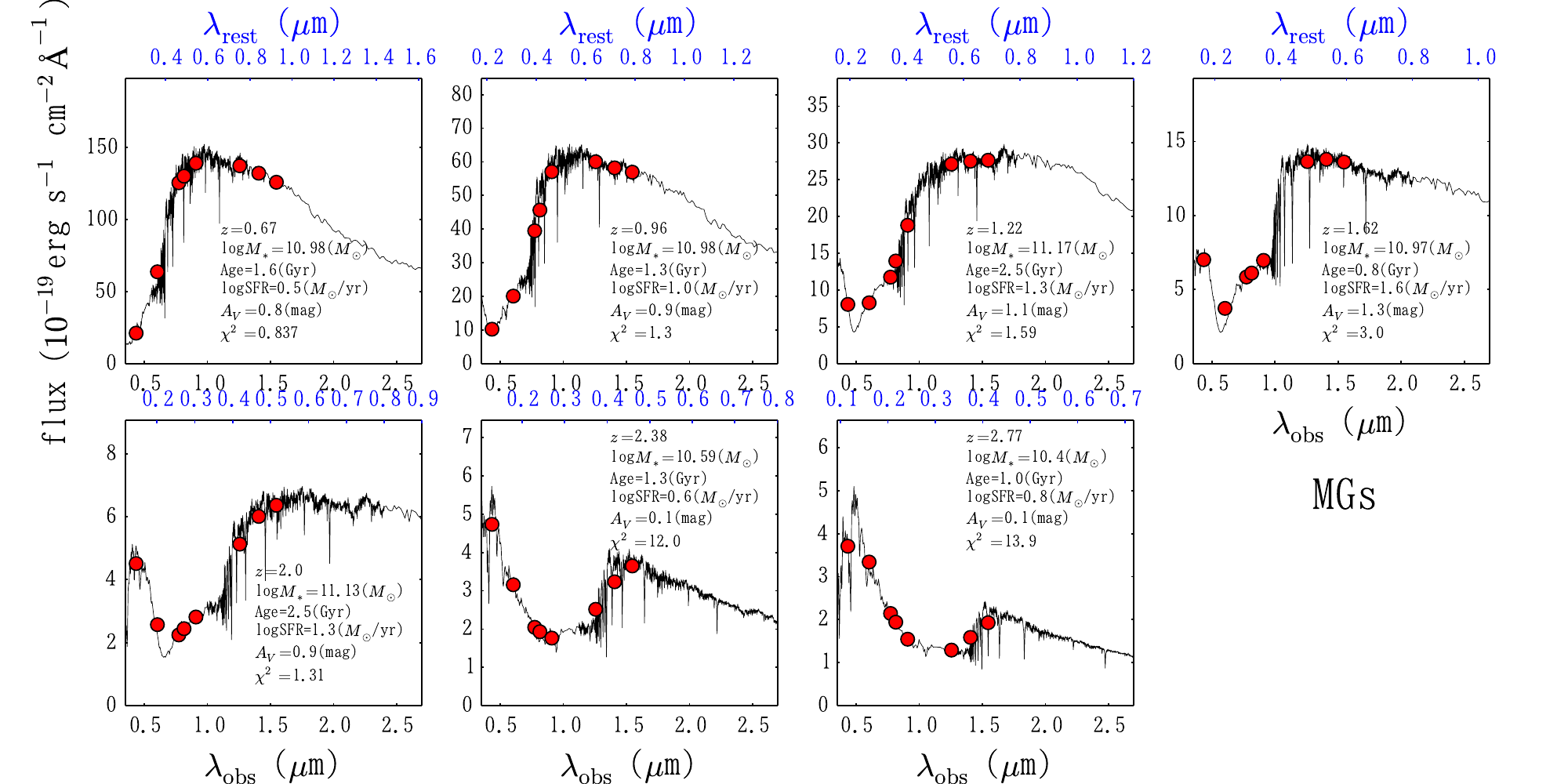}
\caption{
Observed central ($<15$~kpc) fluxes (red circles) for MWs (top) and MGs (bottom).
The best-fit SED templates calculated with FAST are shown with solid lines.
}
\label{fig:flux_all}
\end{center}
\end{figure*}

To study the general properties of the progenitor and to assess the accuracy of the SED fitting, we first integrate the stacked radial profile within the central 15~kpc radius.
By using the stellar population models in Section~\ref{sec:sec_34}, we obtain the SEDs of both samples for each redshift bin and show them in Fig.~\ref{fig:flux_all}.
We find a good agreement for the derived parameters with the median values of each sample, shown in Table~\ref{tb:tb_phys}, within error.
(For the calculation of the error, see the following section.)

\subsection{Radially Resolved SEDs}\label{sec:sec_42}
\begin{figure}
\figurenum{6a}
\begin{center}
\includegraphics[width=9cm, bb=0 0 288 216]{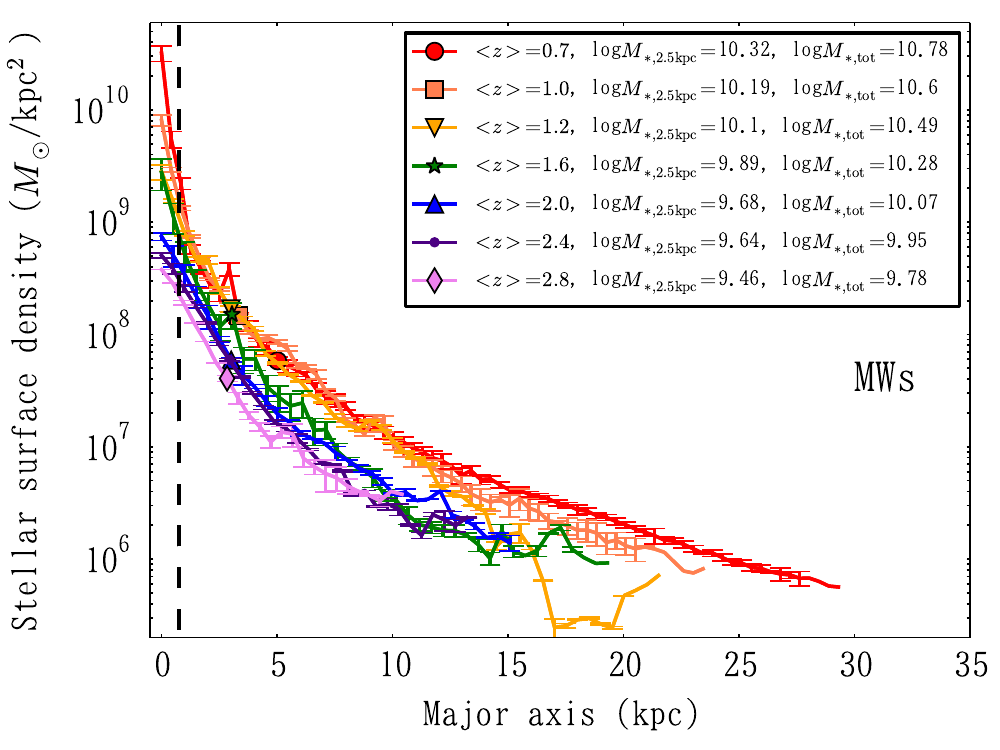}
\includegraphics[width=9cm, bb=0 0 288 216]{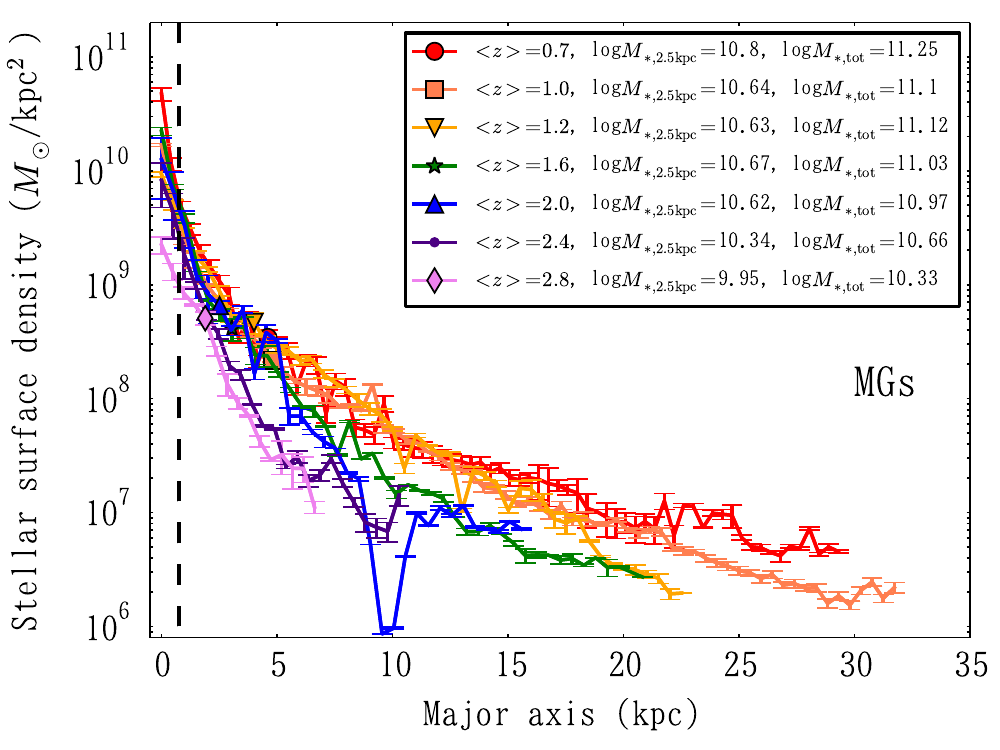}
\caption{
Radial profiles of the stellar surface density of MWs (top) and MGs (bottom).
The position of each symbol represents the half-mass radius, whose value is shown in the caption.
The vertical dashed lines represent the radius that corresponds to the maximum PSF FWHM/2 for the full sample, although the PSF convolution effects are corrected (see the text).
The bar represents the typical error for each radius, which is calculated based on the difference among the three different star-formation histories.
Stellar mass within 2.5~kpc from the center ($M_{*,\mathrm{2.5~kpc}}$) and total stellar mass ($M_{*,\mathrm{tot}}$) for each redshift bin are shown in the caption.
}
\label{fig:fig_mass_mw}
\end{center}
\end{figure}

We obtain the radial SED for the stacked samples.
Using the SED templates prepared in Section~\ref{sec:sec_34}, the radial SED fitting is conducted only for the pixels where the fluxes of F850LP, F125W, and F160W have S/N $>3$.
This is because these bands are sensitive to the Balmer break at high redshift ($z>1.5$).
Instead, we verified this after the Monte Carlo simulation (see Appendix) and found that if the pixel satisfies the criterion, the derived parameters have negligible error compared to, for example, one caused by the assumption in the SED templates (Sec.~\ref{sec:sec_34}).
In Fig.~\ref{fig:fig_mass_mw}, we show the derived surface stellar mass density profiles.
The stellar surface density is obtained down to $\sim10^6~M_\odot$kpc$^{-2}$ at the outer part. 
The error for the stellar mass is estimated from the differences between the star-formation histories, as described in Section~\ref{sec:sec_34}.
The error from the photometric uncertainties is comparatively small ($\sim10^5~M_\odot$), and thus we neglect them hereafter (see Appendix).
 
To conquer the effect of a PSF convolution, we apply the method of Szomoru et al.~(\citeyear{szomoru10}), rather than directly deconvolving the images, to the 1-D stellar mass profiles.
This is because the direct deconvolution could introduce unexpected noise, especially for small size imaging data.
First, we derive the best-fit 1-D S\'ersic profile with GALFIT (Peng et al.~\citeyear{peng10}), taking account for the convolution effect.
Then, we add the residual between the observed and the best-fit profiles to the best-fit $unconvolved$ one.
Although the residual still remains undeconvolved ($\sim1/10$ of the best-fit model), we verify the result by convolving the final ``deconvolved" image with the original PSF kernel for F160W, and then we find that the method reasonably reproduces the observed image, except for outermost parts, which hardly affect our final results.

\begin{figure}
\figurenum{6b}
\begin{center}
\includegraphics[width=9cm,bb=0 0 432 288]{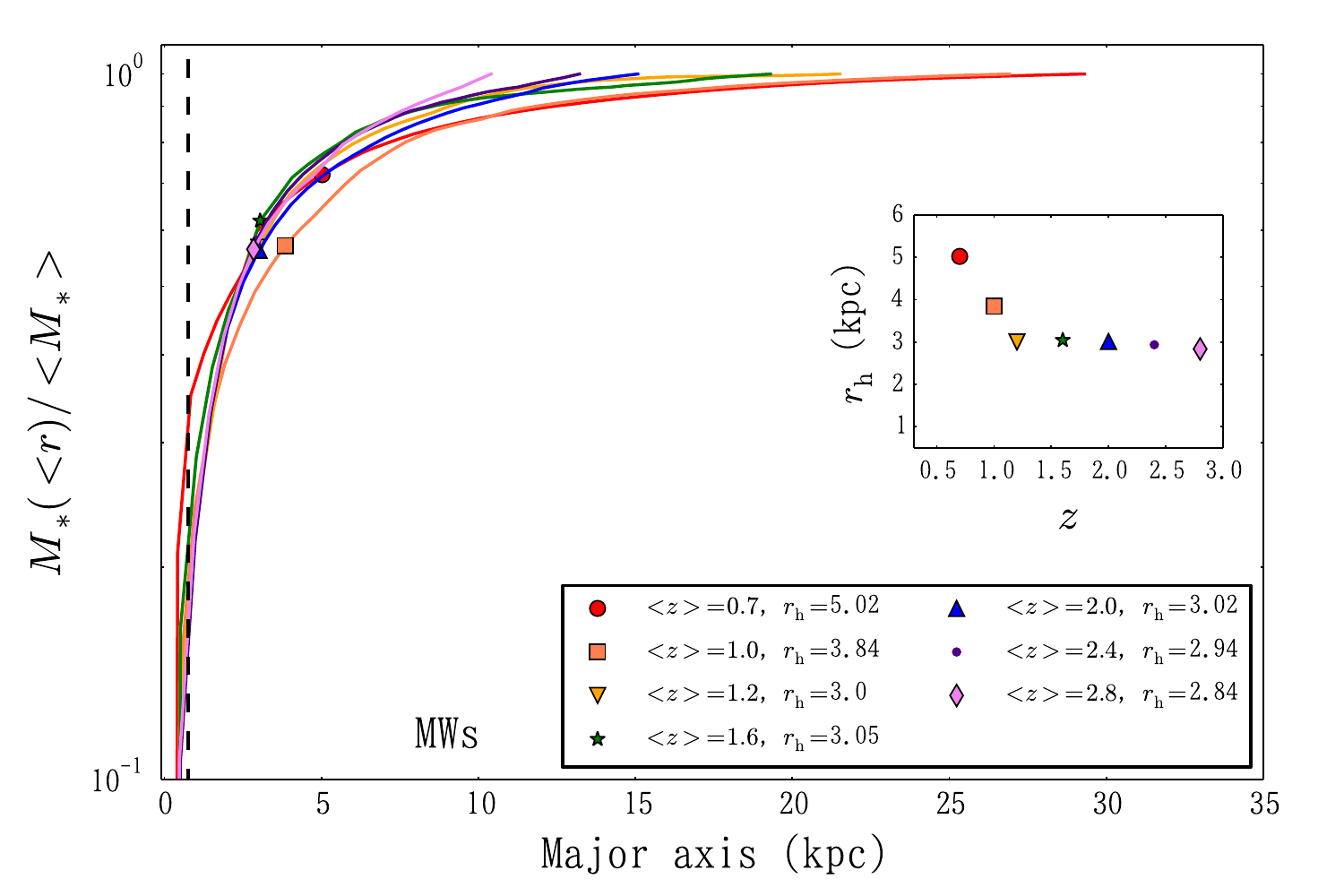}
\includegraphics[width=9cm,bb=0 0 432 288]{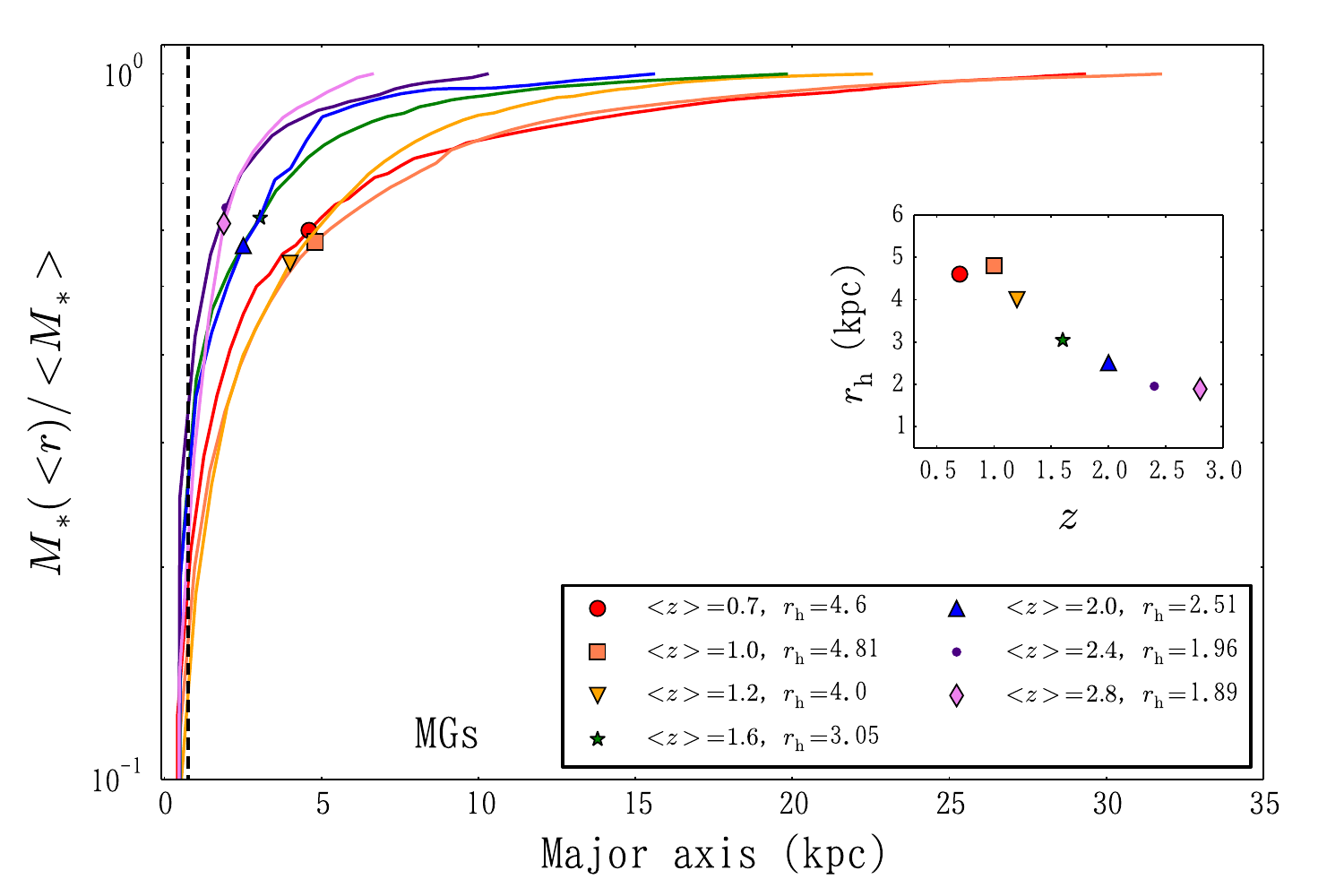}
\caption{
Cumulative stellar mass profiles of MWs (top) and MGs (bottom) for each redshift bin.
Each profile is normalized to the total stellar mass.
The vertical lines and symbols are the same as Fig.~\ref{fig:fig_mass_mw}.
The evolution of the half-mass radius is in the inset.
While the mass profiles of MWs show the self-similar evolution, those of MGs show inside-out evolution, where the  accumulation of stellar mass continues at the outer part of galaxies.
}
\label{fig:fig_mass_mw_cum}
\end{center}
\end{figure}

\begin{figure}
\figurenum{6c}
\begin{center}
\includegraphics[width=9cm,bb=0 0 563 295]{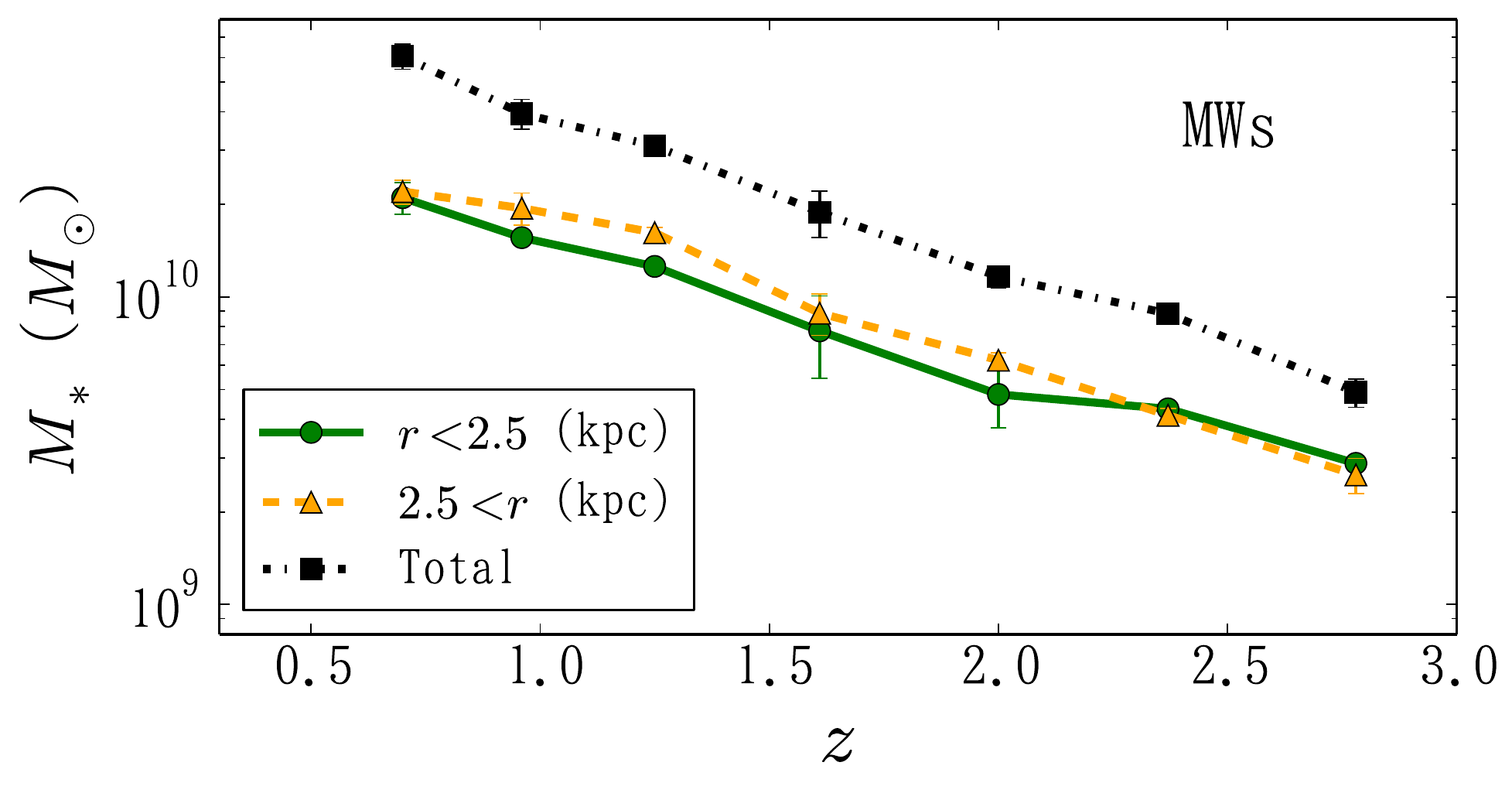}
\includegraphics[width=9cm,bb=0 0 563 295]{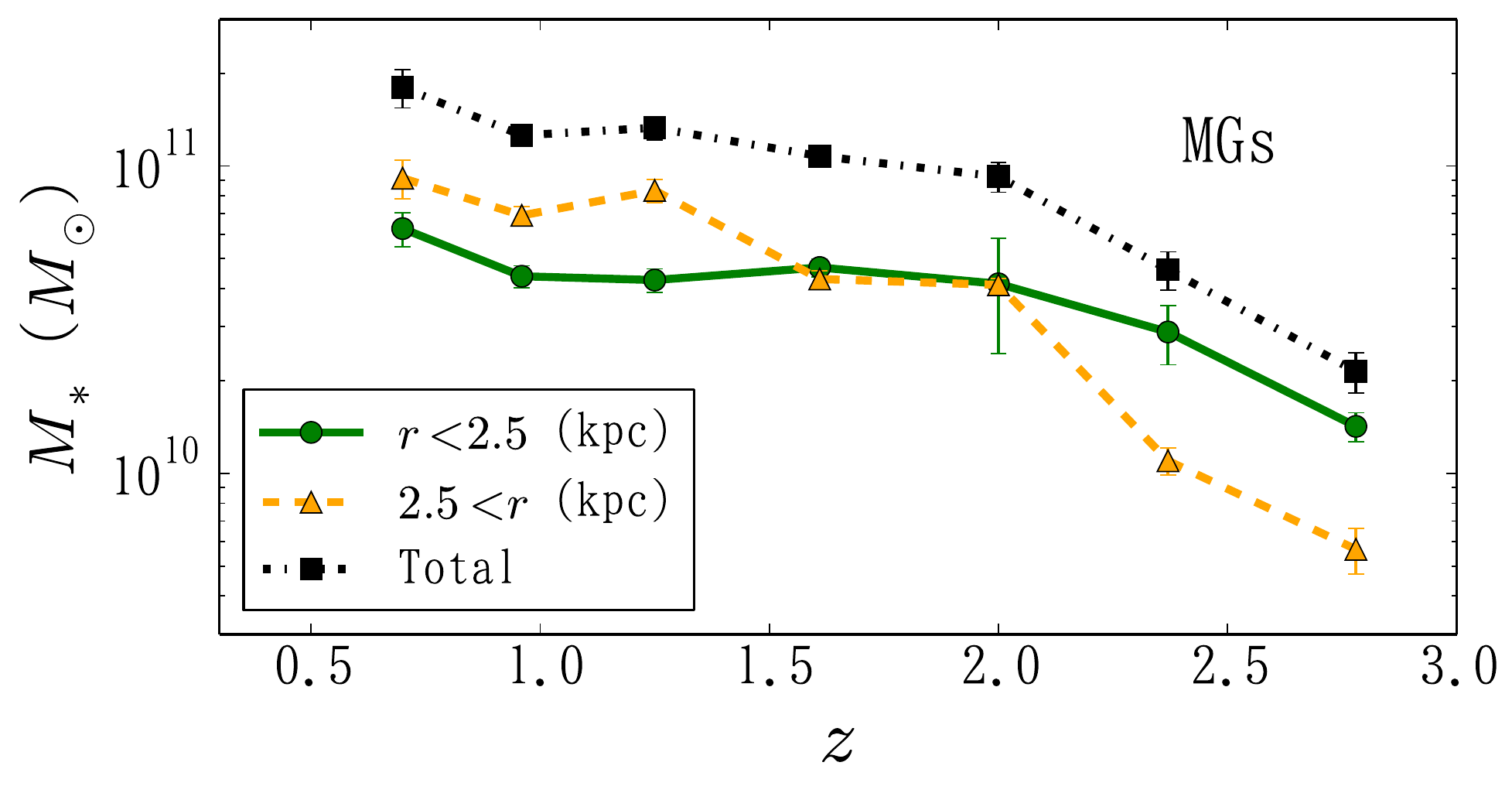}
\caption{
Evolution of the stellar mass for MWs (top) and MGs (bottom); total mass (black squared), mass within $r<2.5$~kpc (green circles), and mass in $r>2.5$~kpc (yellow triangles).
The bar represents the typical error of stellar mass at each region.
We can conjecture two distinct profile evolutions for MWs (self-similar way) and MGs (inside-out way).
}
\label{fig_ms_inout}
\end{center}
\end{figure}

Figure~\ref{fig:fig_mass_mw} shows the deconvolved radial profiles for each redshift bin, where we see distinct differences between the two samples.
The stellar mass of MWs is accumulated at all radii in a similar way (we mention it as a ``self-similar way"), from $z\sim2.8$ to 1.0 with slight surplus increases at the outer part ($>2.5$~kpc) at $z<1$.
The stellar mass of the inner part ($r<2.5$~kpc, hereafter the bulge), grows by a factor of $\sim7$ over the considered redshift range.
\footnote{
We refer to the inner region ($<2.5$~kpc) as ``bulge", because the bulge is expected to dominate the region,
though for more specified definition we need additional analyses, such as the bulge-disk decomposition, which is beyond the present scope.
}
We also show cumulative stellar mass profiles in Fig.~\ref{fig:fig_mass_mw_cum} with non-parametric half-mass-radii.
The only mild change of half-mass-radius ($r_\mathrm{h}\sim2.8$--$5.0$~kpc) over the redshift supports the self-similar evolution.
On the other hand, the half-mass radius of MGs evolves from $\sim1.9$~kpc at $z>2$ to $\sim4.8$~kpc at $z<1$, by a factor of $>2.5$.
The difference of the stellar mass accumulation histories for both samples would become more evident when comparing the stellar masses of the bulge and outer part in Fig.~\ref{fig_ms_inout}.
The comparable growth of both regions for MWs, in the top panel of the figure, supports the self-similar evolution, while we see in the bottom panel the ``inside-out" growth for MGs, where the inner massive bulge has already appeared at $z\sim2.0$ ($M_{*,\mathrm{2.5kpc}}\sim5\times10^{10}M_\odot$), after which the mass increases preferentially at the outer region.

\begin{figure}
\figurenum{7}
\begin{center}
\includegraphics[width=8cm,bb=0 0 302 216]{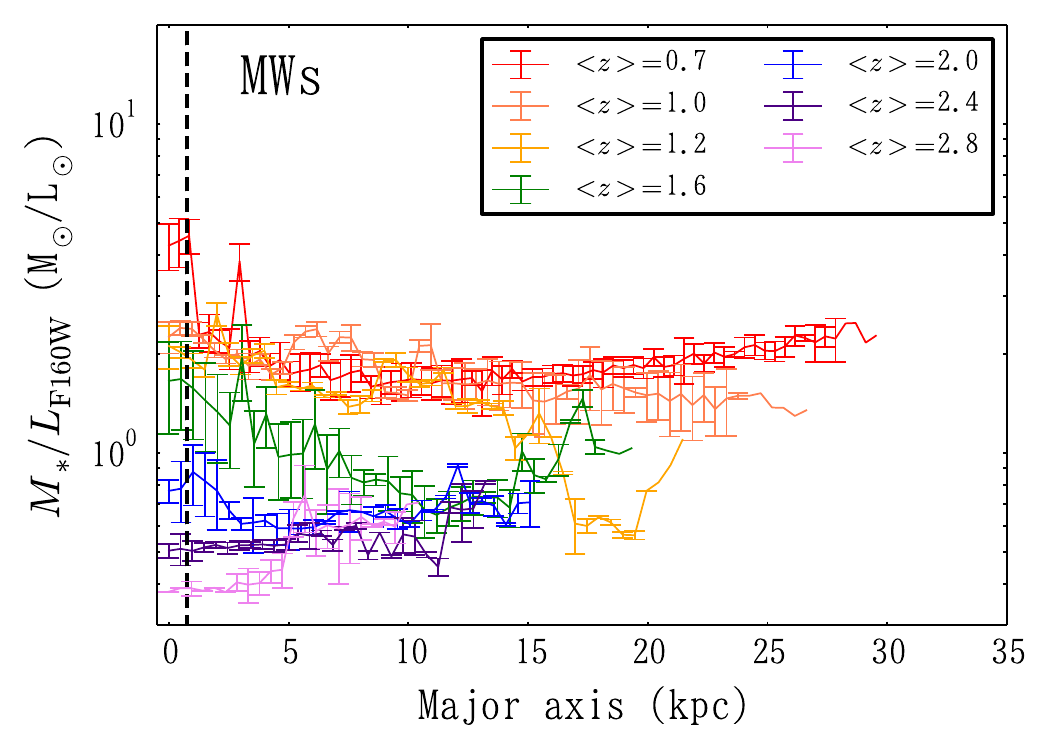}
\includegraphics[width=8cm,bb=0 0 302 216]{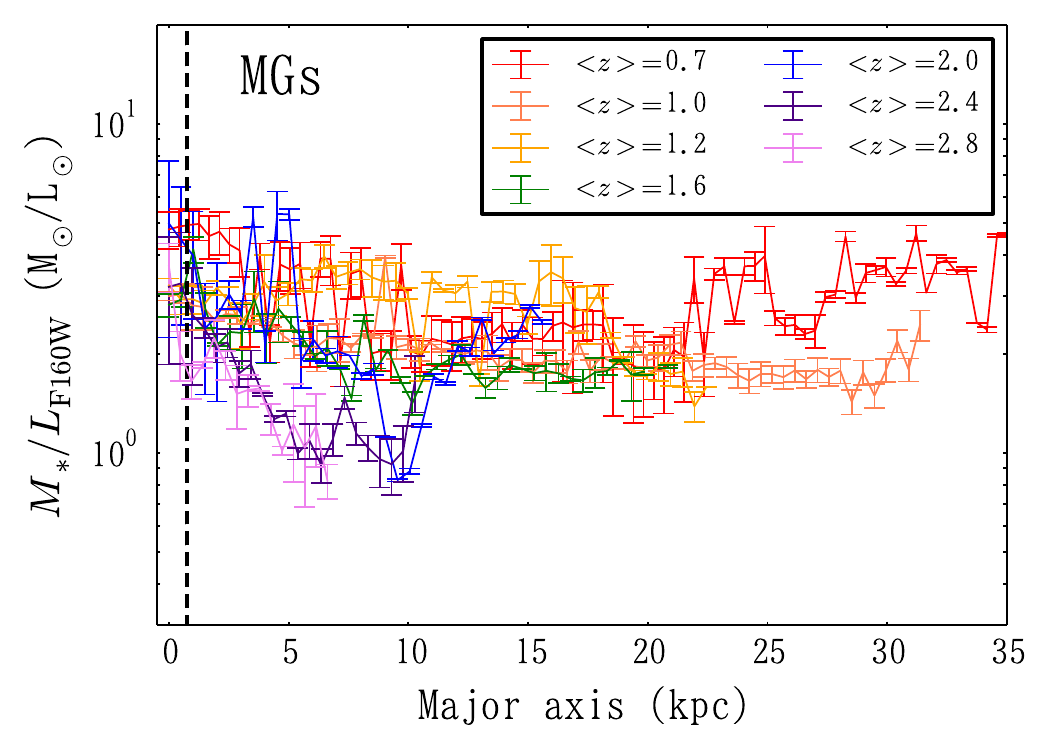}
\caption{
Radial profiles of stellar mass to $H$-band (F160W) light ratio ($M_*/L_\mathrm{F160W}$) of MWs (top) and MGs (bottom) for each redshift.
The bar represents the typical error of stellar mass at each radius.
The vertical lines are same as for Fig.~\ref{fig:fig_mass_mw}.
We see variations of $M_*/L_\mathrm{F160W}$ along galaxy radius, which suggest the necessity of the radially resolved SEDs to derive the radial stellar mass profiles.
}
\label{fig_ml}
\end{center}
\end{figure}

Although the evolution trend of $r_h$ is consistent with those found in the previous studies, the sizes are slightly smaller (for example, $r_e\sim6$~kpc at $z\sim0.5$ for MGs in Patel et al.~\citeyear{patel13}, where $r_e$ is the effective radius calculated by GALFIT).
The disagreement would come from two possible facts; one from the difference between the stellar mass and light profiles, and the other from the difference of the definition of radius (e.g., non-parametric and parametric).
The former offset could be originated by the effect of color gradients (i.e., higher $M/L$ at the central part; e.g., Cappellari et al.~\citeyear{cappellari06}), which would make a radial profile of stellar mass more centrally concentrated.
The radial profiles of stellar mass to the $H$-band light ratio for both populations are shown in Fig.~\ref{fig_ml}.
We find that the slope of $M/L$ varies over the redshift range, which suggests that the mass profile is not correctly reconstructed with only the light profile.
We note that Szomoru et al.~(\citeyear{szomoru13}) also found that half-mass-radii are $\sim25\%$ smaller than half-light radii for the galaxies at $0.5<z<2.5$.
For the latter, the parametric measurement with GALFIT assumes that the outer part of galaxies extends to infinity, which would give larger half-light radius, while the non-parametric method takes account only of really observed stellar mass.
The non-parametric analysis usually suffers from the fact that the outskirts of galaxies at higher redshift could be buried under background noise and missed, while the parametric analysis suffers from the degeneracy of the parameters (S\'ersic index and $r_e$).
We refer the reader to Morishita et al.~(\citeyear{morishita14}) for the detailed discussion of these issues.

\subsection{Rest-Frame Color Gradients}\label{sec:sec_43}
In addition to stellar mass profiles, we investigate rest-frame $UVJ$ colors.
The $UVJ$ colors derived from the best-fit SEDs are used to diagnose high-$z$ quenched and star-forming galaxies (Williams et al.~\citeyear{williams09}; Whitaker et al.~\citeyear{whitaker12}).
We here show the $UVJ$ diagrams and $U-V$ color profiles for MWs and MGs.

\begin{figure*}
\figurenum{8}
\begin{center}
\includegraphics[width=12cm,bb=0 0 690 390]{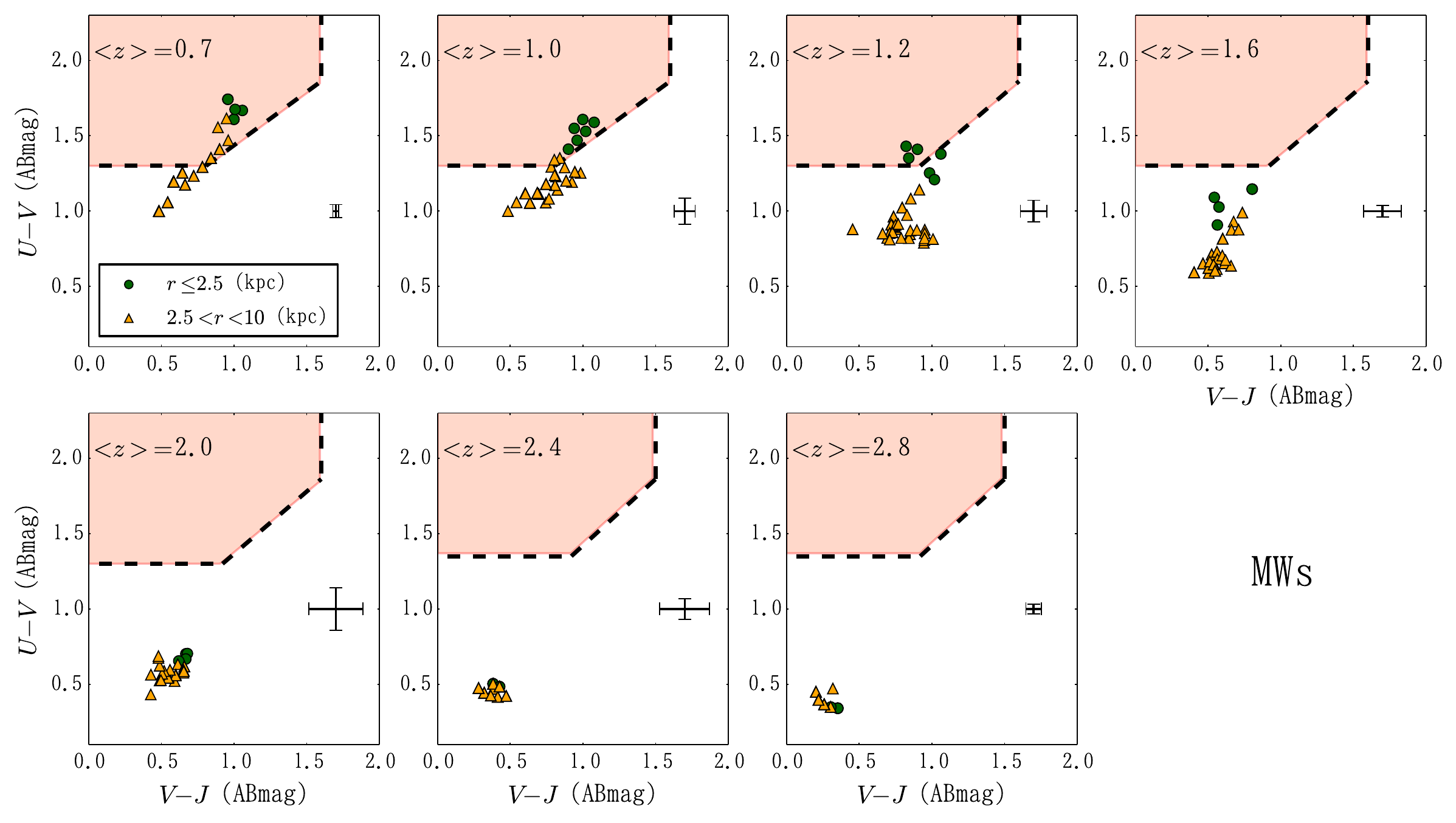}
\includegraphics[width=12cm,bb=0 0 690 390]{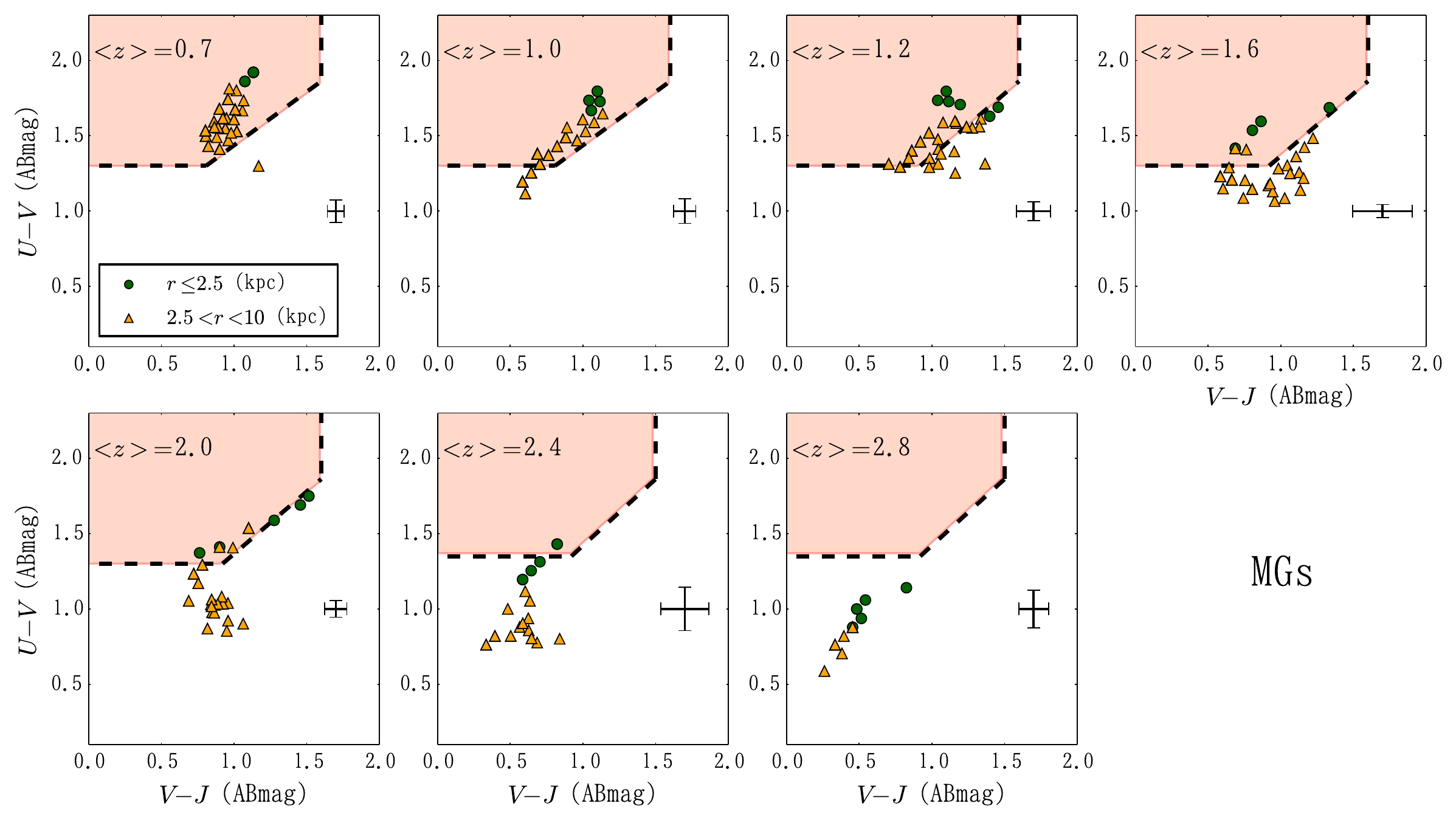}
\caption{
Rest-frame $UVJ$ diagrams for MWs (top) and MGs (bottom).
The symbol represents each pixel, and is distinguished with the distance from the galactic center (green circles for $r\leq2.5$~kpc and yellow triangles for $r>2.5$~kpc).
The dotted lines are the boundary for quiescence (hatched region) defined in Morishita et al.~(\citeyear{morishita14}).
The error bars represent the typical error estimated in Section~\ref{sec:sec_52}.
The evolution of the colors proceeds from the inner to outer regions, which is a common trend for both populations.
}
\label{fig_uvj}
\end{center}
\end{figure*}

First, the $UVJ$ color diagrams of the stacked radial profiles are given in Fig.~\ref{fig_uvj}.
The pixels that enter the top left region, which are bordered with dotted lines, are diagnosed as quiescent, and the others are star-forming.
The criteria were originally defined in Williams et al.~(\citeyear{williams09}) for galaxies at $z<2$.
Morishita et al.~({\citeyear{morishita14}}) refined them for the higher redshift ($z\sim3$) with observed SFRs, and we here adopt refined ones.
(Note that the median specific SFRs in Morishita et al.~({\citeyear{morishita14}}) for QGs at $0.5<z<1.0, 1.0<z<2.0$, and $2.0<z<3.0$ are 0.03, 0.08, and 0.16~per unit Gyr, respectively, while those for star-forming galaxies are 0.57, 1.88, and 2.63~per unit Gyr.)
The symbols are distinguished according to the distance from the galactic center to see the inner ($<2.5$~kpc, or bulge) and outer ($2.5<r<10$~kpc) colors separately, which is inspired by recent findings of the relation between the quiescence and compactness of high-$z$ galaxies (Franx et al.~\citeyear{franx08}; Cheung et al.~\citeyear{cheung12}; Woo et al.~\citeyear{woo14}).
We see that at all redshift bins the inner pixels stand at the reddest frontier, which is in agreement with the general picture of galaxy evolution (Vila-Costas \& Edmunds~\citeyear{vila-costas92}; Wuyts et al.~\citeyear{wuyts12}; Szomoru et al.~\citeyear{szomoru13}).
It is noted that although the dust attenuation makes the galaxy color red, this does not significantly affect the quiescent sample here because dust attenuated galaxies are shifted to just right of the quiescent criteria in Fig.~\ref{fig_uvj} (see arrows in Fig.~\ref{uvj9}; see also Whitaker et al.~\citeyear{whitaker12}).

\begin{figure*}
\figurenum{9}
\begin{center}
\includegraphics[width=8cm,bb=0 0 313 217]{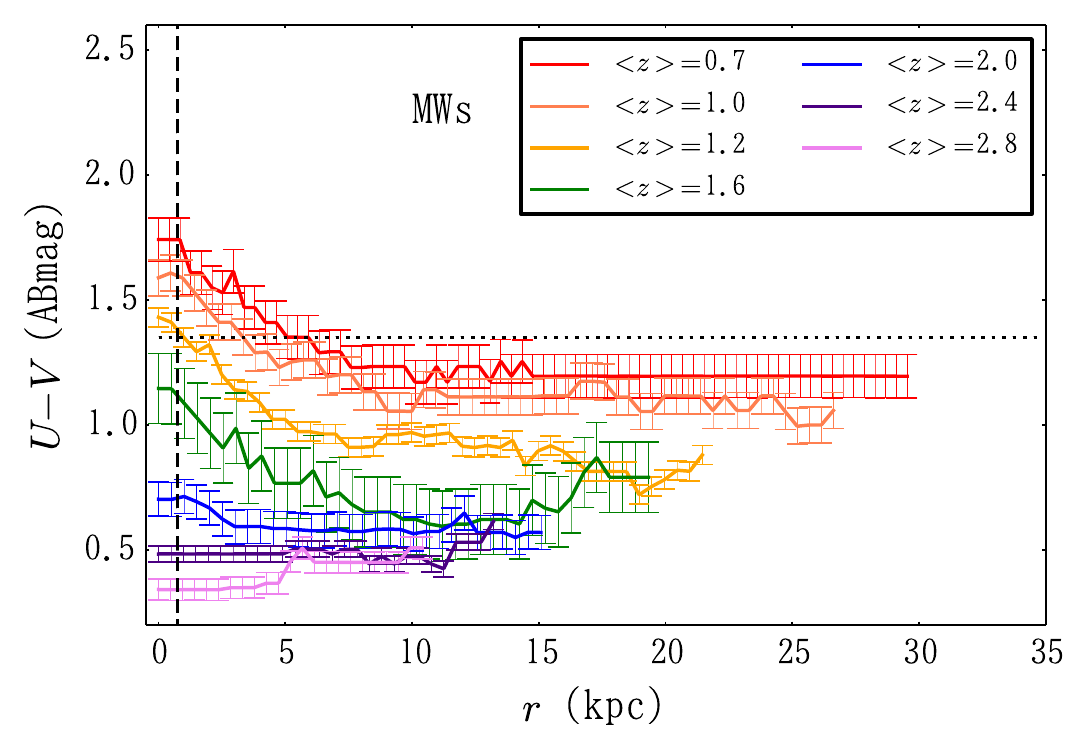}
\includegraphics[width=8cm,bb=0 0 313 217]{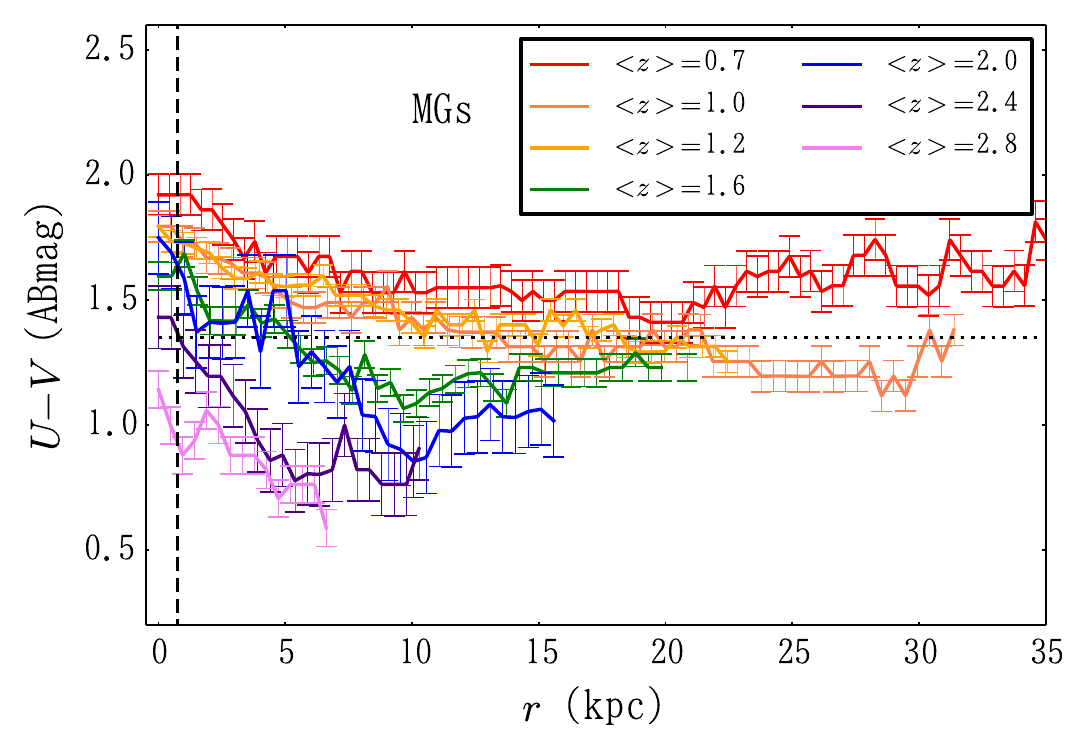}
\caption{
Rest-frame $U-V$ color radial profiles for MWs (top) and MGs (bottom).
The bar represents the typical error.
The same criterion for quiescence as Fig.~\ref{fig_uvj}, $(U-V)=1.35$, is shown with horizontal dotted lines.
}
\label{fig_uv}
\end{center}
\end{figure*}

Figure~\ref{fig_uv} shows the evolution of radial $U-V$ color, which traces the strength of 4000~$\mathrm{\AA}$ and Balmer breaks for both populations, where we can conjecture when and how the quenching starts.
For MWs, we see the bulge start to become redder at $z\sim2$, compared to the outer region.
After that, the bulge seems to quench at $z\sim1.0$, while there is still ongoing star-formation in the outer part, that is, we see the ``coexistence" of quenched bulge and star-forming disk in MWs.
For MGs, on the other hand, quenching seems to start at an earlier epoch ($z\sim2.0$), and star-formation activities seem to stop at $z\sim1.0$ in whole.
We discuss these findings in the following section along with the result of the radial stellar mass profiles.

\section{DISCUSSION}\label{sec:sec_5}
In the last two sections we stacked the multi-band 1-D light profiles of two populations (MWs and MGs) in each redshift bin to obtain the composite radial profiles, and then derived the radial SED to see the transitions of the radial mass profiles over the redshift range.
With the radially resolved SED profiles, we are now able to discuss not only where the stellar mass growth happens, but also what mechanisms facilitate the evolution.
Furthermore, the stacked profile allows us, for the first time, to evaluate the dispersion of radial profiles of individual galaxies around the median profile at a given redshift, as an indicator of morphological variety.
Examining this variety gives us a clue to the evolution of galaxy morphology, in the context of the cosmic star-formation history over the redshift and manifestation of the Hubble sequence.

\subsection{Inside-out Mass Growth Versus Self-similar Mass Growth}\label{sec:sec_51}
As shown in Section~\ref{sec:sec_42}, the stellar mass profiles of MGs and MWs evolve in different ways.
The former evolves in an $inside$-$out$ way, while the latter evolves in a $self$-$similar$ way.
In the inside-out scenario, it is claimed that massive naked bulges appeared at $z\sim2$ (or even earlier; Nelson et al.~\citeyear{nelson14}), and then it gains the rest of mass at the outer part by accreting less massive satellite galaxies, as shown in early observation studies (e.g., Trujillo et al.~\citeyear{trujillo12}).
We clearly see in the bottom panels of Figs.~\ref{fig:fig_mass_mw} and \ref{fig:fig_mass_mw_cum} that the massive bulge of the MGs (log$M_{*,2.5\mathrm{kpc}}>10.5$) was formed at $z\sim2.4$, and then the mass growth mostly occurred at the outer part ($>2.5$~kpc).
This can be also found when we compare the mass of the bulge to that of the outer disk in the bottom panel of Fig.~\ref{fig_ms_inout}, where we find that the former does not evolve after $z\sim2$, while the latter continuously evolves over the entire redshift range.
Furthermore, the observed growth of the bulge and total stellar mass suggests that more than $\sim75\%$ of the stellar mass is accumulated at the outer envelope of galaxies after $z\sim2$.
Our findings for MGs support previous studies with light profiles (van Dokkum et al.~\citeyear{vandokkum10}; Patel et al.~\citeyear{patel13}).

The inside-out, or ``two-phase" (Naab et al.~\citeyear{naab09}; Oser et al.~\citeyear{oser10}), evolution scenario explains the previously observed size growth of $r_e$ by a factor of $\sim2$-5 (Bezanson et al.~\citeyear{bezanson09}; van Dokkum et al.~\citeyear{vandokkum10}), although the scenario is still in dispute (e.g., Carollo et al.~\citeyear{carollo13}; Poggianti et al.~\citeyear{poggianti13}; Riechers et al.~\citeyear{riechers13}).
According to the scenario, both effective radius and S\'ersic index are expected to evolve (Morishita et al.~\citeyear{morishita14}), whereas in this work we only measure the non-parametric half-mass radius to avoid the parametric degeneracy.
The formation of bulges at $z>2$, on the other hand, would be difficult to expect in such a simplified and continuous scenario, and at the higher redshift violent dynamical mechanisms, such as major merger and clump migration, might trigger the formation of bulges (Barro et al.~\citeyear{barro14}; Dekel et al.~\citeyear{dekel14}).

On the other hand, we find the mass profiles of MWs evolving in self-similar way, where the mass accumulation at the inner and outer parts are comparable (see the top panel of Fig.~\ref{fig_ms_inout}).
This similarity is a consequence of the continuous bulge growth even at $z<1.0$, while MGs show little or no bulge evolution after $z\sim2$.
The result is in good agreement with vD13, while there are other inconsistent observational evidences for two- or three-phase formation scenarios for the MW where the bulge and thick disk evolve first, and then the thin disk enlarges the galaxy size (Toyouchi \& Chiba~\citeyear{toyouchi14}; see also Chiappini~\citeyear{chiappini01}).
Although we could not elucidate the true evolution process only by observing stellar mass profiles (because they do not tell us specific dynamical processes, e.g., stars radially mixing; Sellwood \& Binney~\citeyear{sellwood02}), merging of galaxies and AGN feedback (Fan et al.~\citeyear{fan08}), in the following give a possible explanation for the self-similar evolution, on receiving the results of rest-frame colors.

\subsection{Rest-Frame Color Profile as a Diagnosis of Quenching}\label{sec:sec_52}
In this subsection we investigate the shutdown of star-formation activities (quenching) for both populations by using the stellar mass and the rest-frame color profiles derived in Section~\ref{sec:sec_4}.
The $U-V$ colors (Figs.~\ref{fig_uvj} and \ref{fig_uv}) tell us how the star-forming region becomes quiescent, namely, ``how the quenching proceeds."
We see that both populations quench the star-formation activity from inner to outer radii.
The trend is clear in Fig.~\ref{fig_uv}, where the inner regions keep reddest at any redshift bins. 

\begin{figure*}
\figurenum{10a}
\begin{center}
\includegraphics[width=6cm,bb=0 0 204 209]{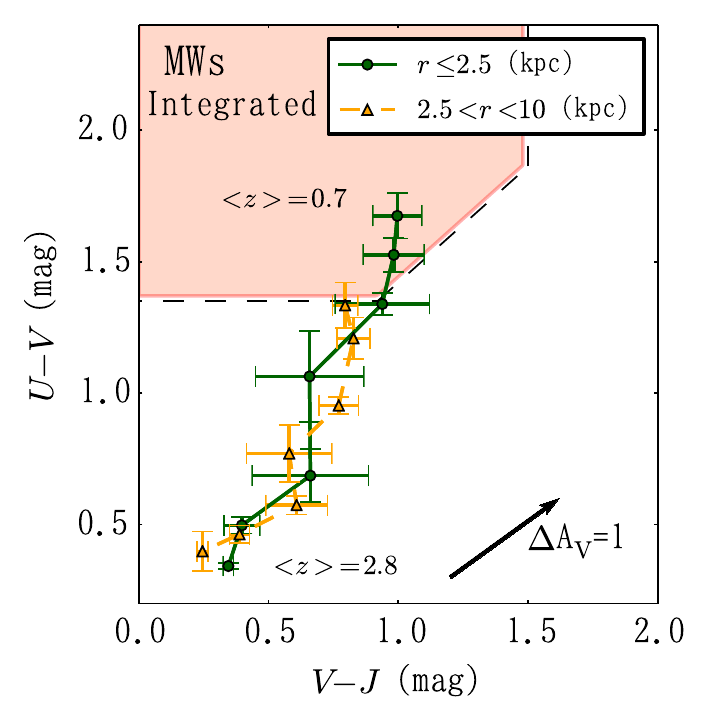}
\includegraphics[width=6cm,bb=0 0 204 209]{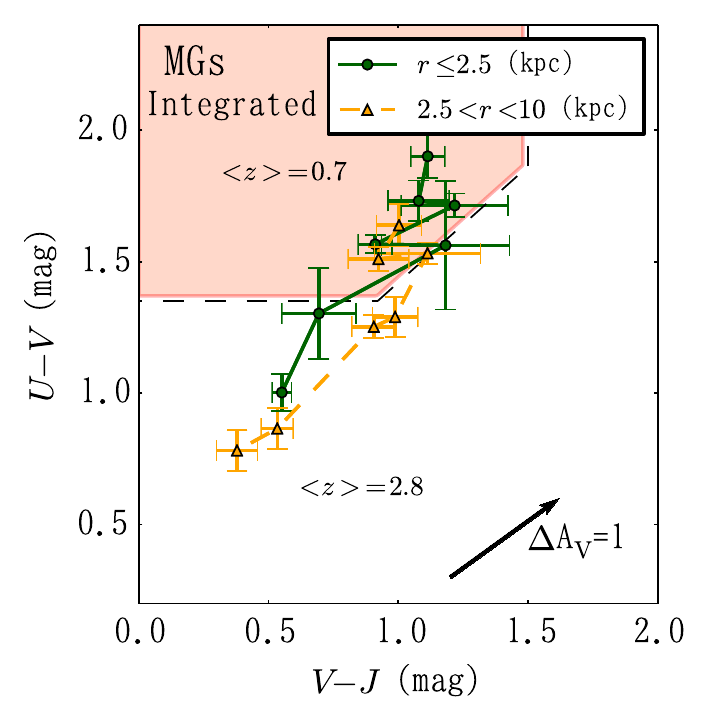}
\caption{
Evolutions of rest-frame $UVJ$ colors for MWs (left) and MGs (right).
The symbols represent the integrated colors of the inner region ($r\leq2.5$~kpc, green circles) and outer region ($2.5<r<10$~kpc, yellow triangles) at each redshift, from the bottom left ($z\sim2.8$) to the top right ($z\sim0.7$).
The bar represents the typical error, which is estimated in Section~\ref{sec:sec_52}, for each redshift.
The boundary with the dashed lines and hatched region are same as for Fig.~\ref{fig_uvj}.
The effect of dust attenuation ($\Delta A_V=1.0$~mag) is shown with an arrow.
}
\label{uvj9}
\end{center}
\end{figure*}

\begin{figure*}
\figurenum{10b}
\begin{center}
\includegraphics[width=6cm,bb=0 0 276 299]{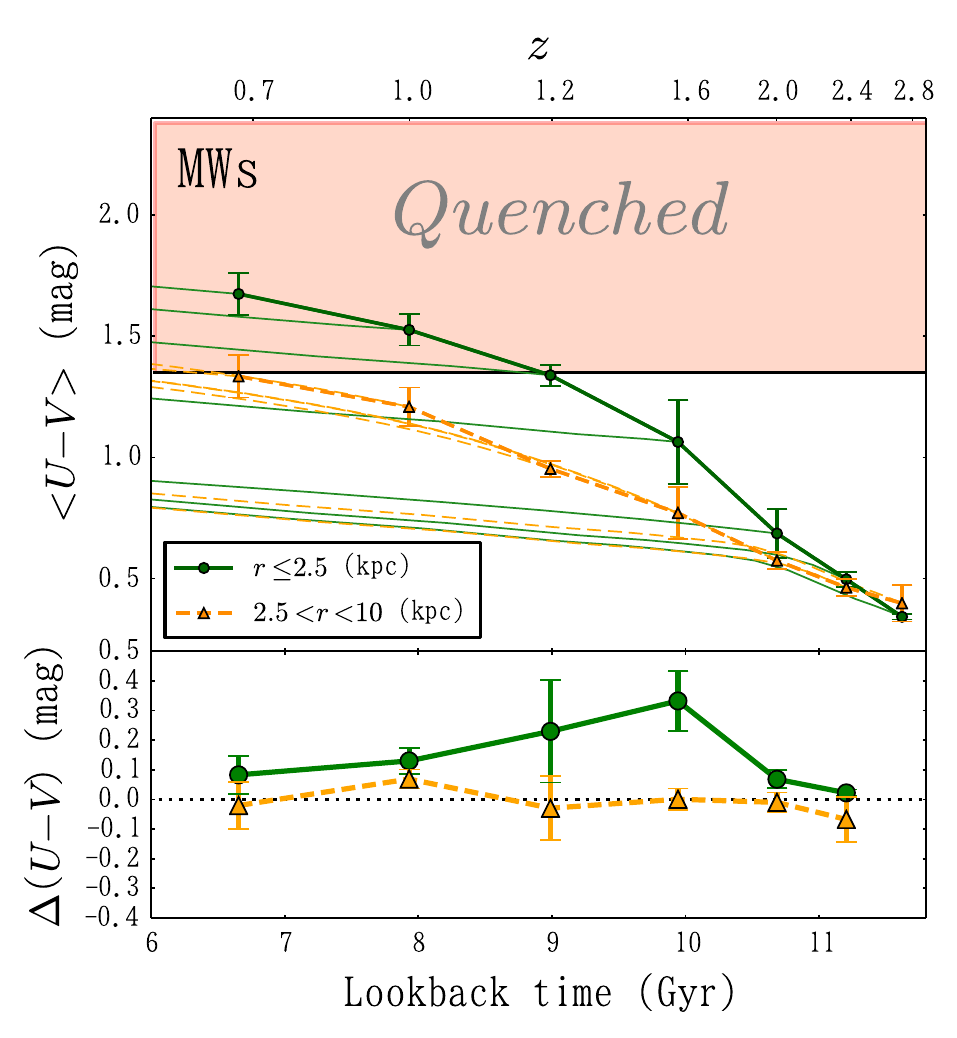}
\includegraphics[width=6cm,bb=0 0 276 299]{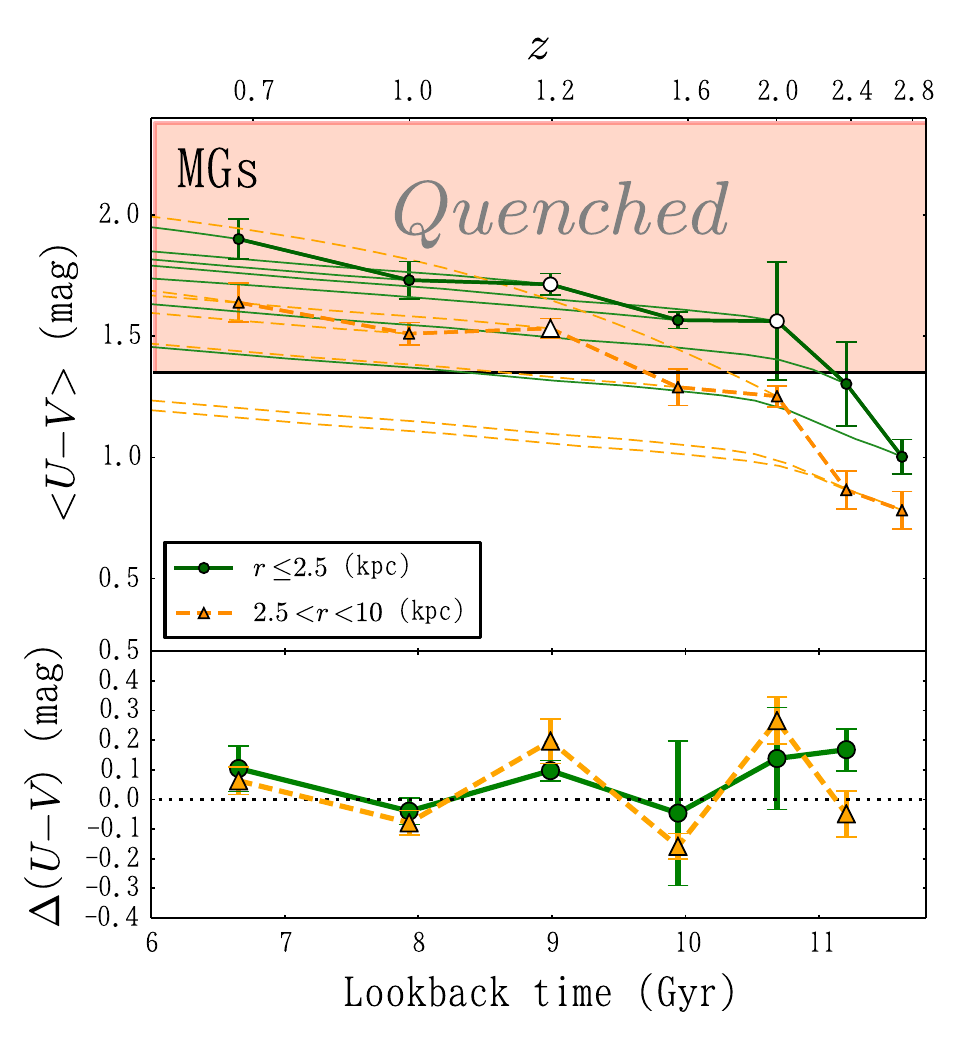}
\caption{
Top: evolution of the rest-frame $U-V$ colors of the inner ($r<2.5$~kpc; green circles) and outer ($2.5<r<10$~kpc; yellow triangles) regions of MWs (left) and MGs (right).
The error bar represents the typical error for $U-V$ at each redshift.
The hatched quench region is set with $(U-V)>1.35$~mag based on Fig.~\ref{uvj9}.
The data points which locate out of the quiescent hatched region in Fig.~\ref{uvj9} within the error, whereas $(U-V)>1.35$~mag are shown with open symbols to distinguish them from the quenched sample.
The model predictions of $U-V$ color (thin lines) are depicted with the best-fit SED profiles at each redshift and the population synthesis model of GALAXEV, to see when each region becomes ``unusually" red or quenched.
Bottom: Comparison of the observed and model $U-V$ colors at each redshift, where we define $\Delta (U-V)\equiv (U-V)_{obs}-(U-V)_{model}$.
The errors are estimated in the same manner for the top panel.
We see the excess of $(U-V)_{obs}$ in the inner regions ($r<2.5~$kpc) at $z\sim1.6$ and $z\sim2.8$ for MWs and MGs, respectively, but no excess are seen in the outer regions for either sample.
}
\label{uv9}
\end{center}
\end{figure*}

For further investigation, we show the color evolution of bulges and outer regions of MWs and MGs in Fig.~\ref{uvj9}.
The $UVJ$ criterion is same as in Fig.~\ref{fig_uvj}, and the rest-frame colors for both samples are derived after summing up the pixels in the bulge and outer region.
Shown in the top panels of Fig.~\ref{uv9} is the evolution of $U-V$ color, where we set the ``quenched" region (hatched) as $(U-V)>1.35$.
The error is estimated in the same manner for stellar mass (see Section~\ref{sec:sec_34}).
The trend is similar to what we see in Fig.~\ref{fig_uvj}.
Although the definition of ``quench" (or quiescence) is based on both $U-V$ and $V-J$ colors, we verify in Fig.~\ref{uvj9} that most of the data points that have $(U-V)>1.35$ are in the quiescent region of the $UVJ$ colors, and therefore we hereafter adopt the $U-V$ color as an indicator of quenching.
The data points of $(U - V) > 1.35$~mag located outside the quiescent hatched region in Fig.~\ref{uvj9}, but within the error, are shown with open symbols to distinguish them from the quenched sample.
In the figure, we see the similarity of the bulge and outer colors at the highest redshift, for each sample.
As redshift decreases, on the other hand, we find that the bulge becomes redder than the outer part (at $z\sim1.6$ for MWs and $z\sim2.4$ for MGs).
We conjecture two possibilities for bimodality of the colors; 
one is reddening of the bulge through quenching, and another is increase of star-formation activity at the outer part, which leads to bluer color.

To investigate this, we calculate for each epoch the expected color at the next epoch by using the best-fit SED template and population synthesis model, and then compare it with the actually observed color at the next epoch.
The SED model is prepared with the best-fit parameters derived in the same manner as the previous section, and its evolution is calculated with GALAXEV by following the derived SED parameters (SFR, $\tau$ and age) at each redshift to the next redshift.
Shown in the top panel of Fig.~\ref{uv9}, the thin line is the expected evolutions of $U-V$ color for each redshift.
Then, we obtain the model $U-V$ color, $(U-V)_{model}$, at the $i$th redshift, $z_i$, by using the one expected from the $(i-1)$th redshift to compare with the observed $U-V$ color, $(U-V)_{obs}$, at $z_{i}$.
Since GALAXEV does not include quenching functions (e.g., gas stripping or AGN feedback), the model and observed quantities should be matched unless such quenching mechanisms work.
The comparisons of the colors for both populations are depicted in the bottom panels of Fig.~\ref{uv9}, with $\Delta (U-V) \equiv (U-V)_{obs} - (U-V)_{model}$.
From the figure, we clearly see that the bulges of MWs and MGs become unusually red at $z\sim1.6$ and $z\sim2.4$ (over a timescale of $\sim1$~Gyr), respectively, while the outer parts of both populations show neither red nor blue excess of color.
Therefore, we conclude that the bimodality is caused by the shutdown of star-formation activities in the inner regions (hereafter referred to as ``bulge quenching"), while the outer parts evolved as expected by the best-fit SED models.

For the bulge quenching, we mention several possible processes suggested by recent theoretical and observation studies.
The first is the termination of the cold gas supply by virial shock at the halo radius.
The halo masses, $M_\mathrm{h}$, of MWs and MGs at the quenching epoch are estimated to be $M_h \sim10^{12}M_\odot$ at $z\sim1.6$ and $M_\mathrm{h}\sim 3\times10^{12}~M_\odot$ at $z\sim2.4$, respectively, based on the result of Behroozi et al.~(\citeyear{behroozi13_770}).
Although the quenched halo masses of both populations are not identical, this is consistent with, for example, Dekel \& Birnboim~(\citeyear{dekel06}), where cold gas continues to flow into the central part of massive halo at higher redshift ($M_\mathrm{h}>10^{12}M_\odot$ at $z\sim2$).
At lower redshift, on the other hand, all of it is shock-heated, even in the system of lower halo mass ($M_\mathrm{h}\sim 7\times10^{11} M_\odot$).
The termination of gas flow into the central region is also explained by the fact that gas with smaller angular momentum preferentially falls into the central region while one with larger momentum, which accretes at a later epoch, cannot fall and thus remains at the outer radius.
The second possibility is the one called ``morphological quenching" (Martig et al.~\citeyear{martig09}), where the formation of the bulge stabilizes the gas kinematics and prevents the system from forming stars.
Several recent observations with IFUs have found that the bulges of high-$z$ massive galaxies are indeed stabilized with high Toomre $Q$-parameter (e.g., Genzel et al.~\citeyear{genzel14}).
The findings of the correlation between the quiescence and morphological compactness also support this scenario (Franx et al.~\citeyear{franx08}; Cheung et al.~\citeyear{cheung12}; Woo et al.~\citeyear{woo14}).
The last possible process is the radiative energy feedback from low mass stars in a compact system, which effectively prevents the formation of H$_2$ molecule and stops star-formation (Kajisawa et al.~\citeyear{kajisawa15}).

In addition, the stellar mass accumulations for both populations (Section~\ref{sec:sec_42}) can be investigated with the result of rest-frame colors (Section~\ref{sec:sec_43}). 
As shown in Fig.~\ref{fig_ms_inout}, the bulge of MWs continuously evolves, even after its quenching epoch at $z\sim1.0$.
This suggests that there should be a mechanism that transports a huge amount of stars ($\sim5\times10^9 M_\odot$) into the bulge from the outer disk and/or external systems in 1~Gyr or so.
The bulge of MGs is rapidly formed to be $\sim4\times10^{10} M_\odot$ at $z>2.0$, where the SFR of MGs peaks (see the following section), and after that we see no evolution.
The outer part of MGs, on the other hand, still continues to form stars ($\sim5\times10^{10}M_\odot$) after $z<1.5$, where the star-formation seems to stop (Figs.~\ref{fig_uvj} and \ref{fig_uv}).
The continuous growth of the outer envelope is consistently explained with $\sim2$ merger events of the gas poor satellite galaxy with $\sim3\times10^{10}M_\odot$, which is in good agreement with recent observation studies (e.g., Ferreras et al.~\citeyear{ferreras14}).
The gas poor satellite merger would enlarge the S\'ersic index of galaxy, as was shown in our previous study of massive QGs (Morishita et al.~\citeyear{morishita14}).

\subsection{Morphological Variation and Appearance of the Hubble Sequence}\label{sec:sec_53}
\begin{figure*}
\figurenum{11}
\begin{center}
\includegraphics[width=12.cm,bb=0 0 295 121]{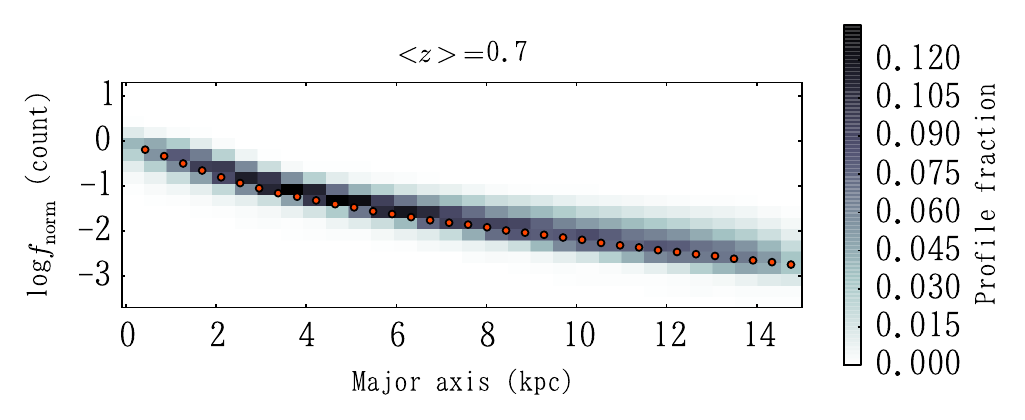}
\includegraphics[width=12cm,bb=0 0 206 82]{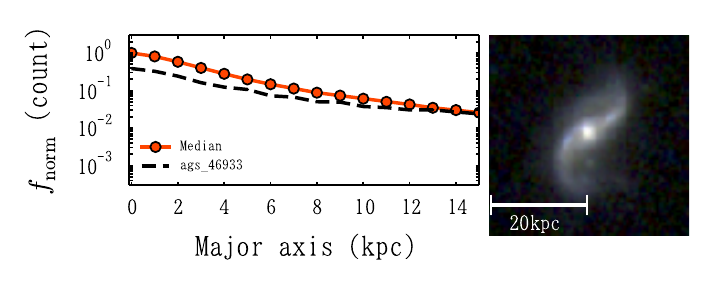}
\includegraphics[width=12cm,bb=0 0 206 82]{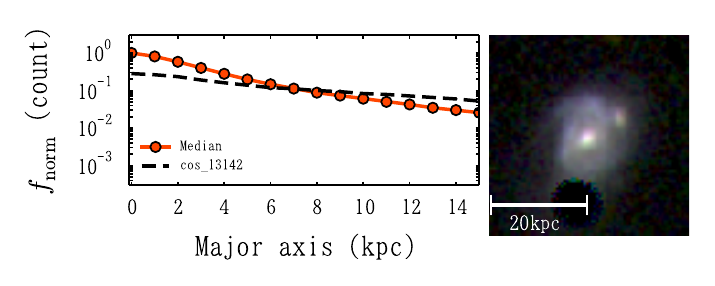}
\caption{
Examples of the radial light profile for MWs at $z\sim0.7$.
Top: gray scale represents the fraction of radial light (F160W) profiles of the galaxies used for stacking.
The median profile is shown as a thick solid line with circles.
The flux is normalized at the center of the median profiles.
Middle and bottom panels: Radial light profiles of randomly selected samples (dotted lines), compared with the median profiles (solid line; same as in the top panel). 
The rgb color pictures composited with F814W (blue), F125W (green), and F160W (red) bands are shown in each panel with the physical size.
The black blot in the bottom rgb image is the masked region.
The pictures of all samples are available at \href{http://www.astr.tohoku.ac.jp/~mtakahiro/mori15apj/mori_15_rad.pdf}{http://www.astr.tohoku.ac.jp/$\sim$mtakahiro/mori15apj/mori\_15\_rad.pdf}.
}
\label{fig_rad2}
\end{center}
\end{figure*}

We here study the varieties of individual profiles from the median stacked profiles and their evolution (see also Kajisawa \& Yamada~\citeyear{kajisawa01}).
In the top panel of Fig.~\ref{fig_rad2} we superpose all radial luminosity profiles of galaxies in $z\sim0.7$ bin, and compare them with the median radial profile
\footnote{The whole samples are available at \href{http://www.astr.tohoku.ac.jp/~mtakahiro/mori15apj/mori_15_rad.pdf}{http://www.astr.tohoku.ac.jp/$\sim$mtakahiro/mori15apj/mori\_15\_rad.pdf}}
.
The median and individual profiles here are the luminosity in F160W-band, since we did not obtain the stellar mass profile for each sample.
The individual light profiles are reduced in the same manner for the radial SED, including a mask for background galaxies and redshift/stellar mass correction of Eq.~(\ref{eq_k}).

\begin{figure*}
\figurenum{12}
\begin{center}
\includegraphics[width=\textwidth,bb=43 47 1695 900]{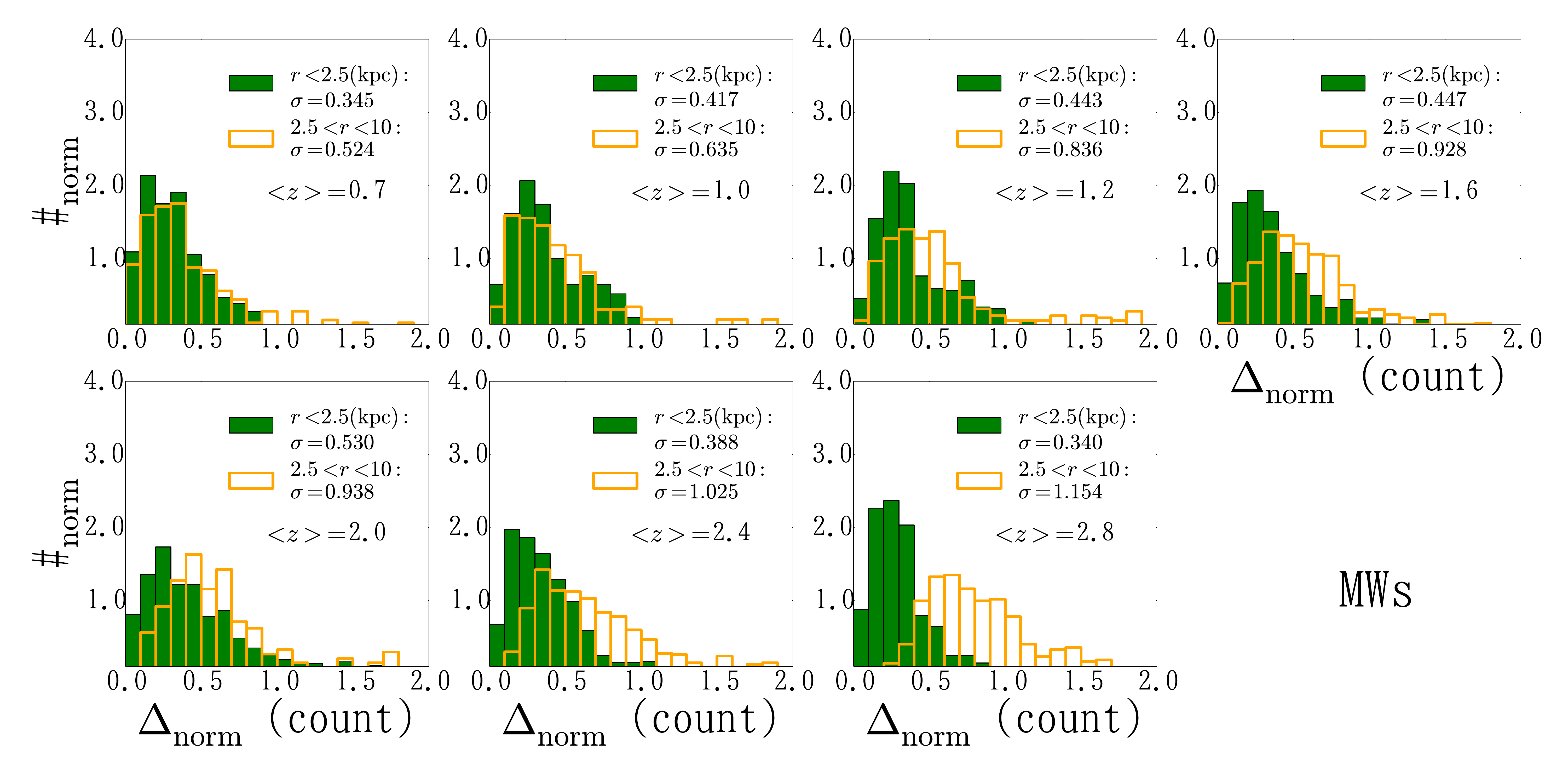}
\includegraphics[width=\textwidth,bb=43 47 1695 900]{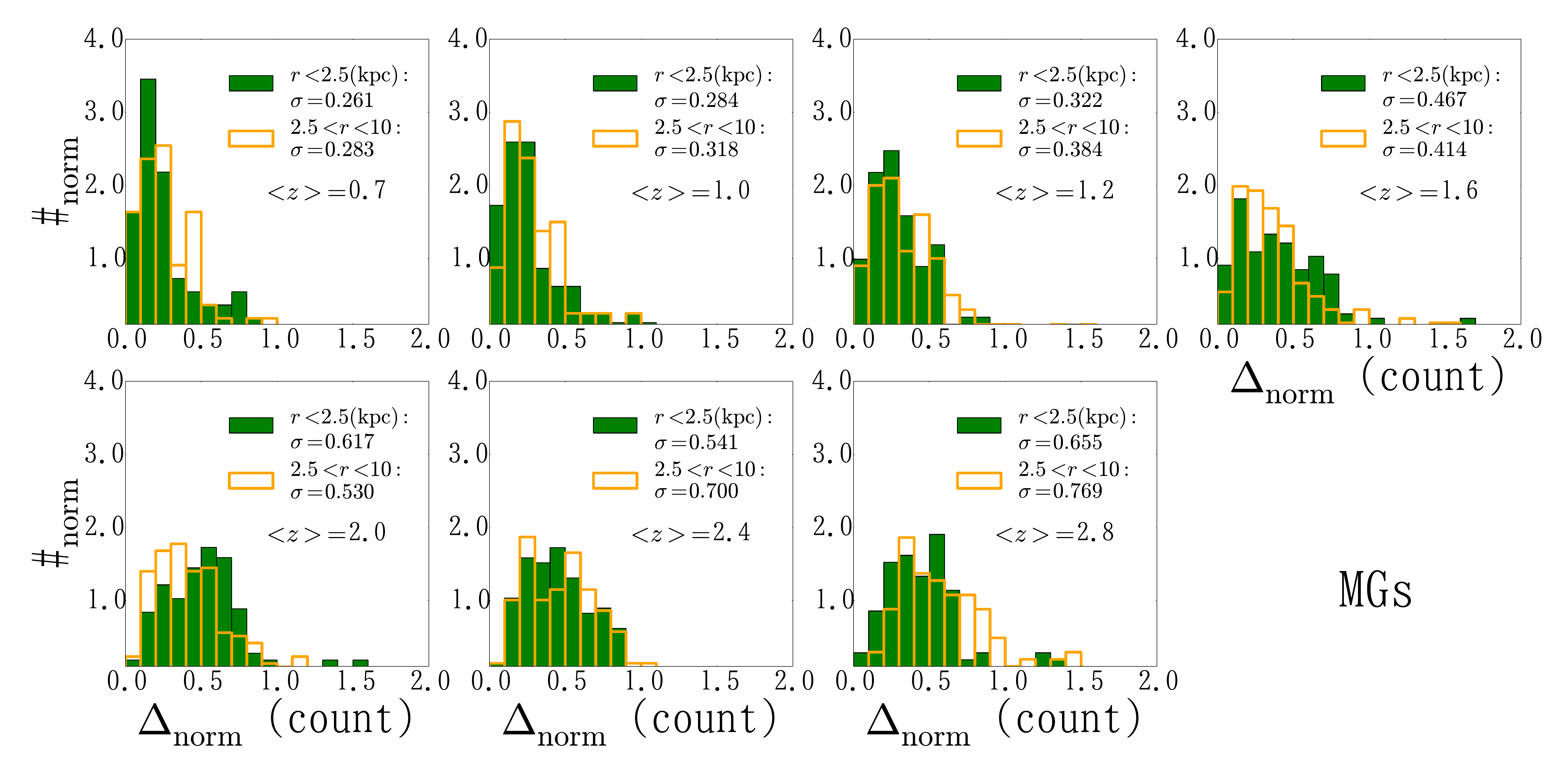}
\caption{
Histograms of the radial variations, $\Delta_\mathrm{norm}$ (see the main text for the definition), for MWs (top) and MGs (bottom) of $r<2.5$~kpc (filled green histograms) and $2.5<r<10$~kpc (unfilled yellow) for each redshift panel.
The dispersion ($\sigma$) of distribution, which represents the variety of galaxy morphology, is shown in each panel. 
}
\label{fig_rad4_each}
\end{center}
\end{figure*}

We define the variety of the galaxy profile by evaluating the distribution of residuals as,
\begin{equation}
\Delta_{\mathrm{norm},x}  = {1\over S_x} \sum^x_{i} | {{f_{i,\mathrm{median}} - f_{i,\mathrm{obs}}}\over f_{i,\mathrm{median}}}|,
\end{equation}
where we set $x<2.5$~kpc and $2.5<x<10$~kpc in the following to see the radial dependence of the variety.
$S_x$ is the total pixel within the range of $x$.
In Fig.~\ref{fig_rad4_each} we show the distribution of $\Delta_{\mathrm{norm},x}$ for each redshift bin.
Then we evaluate the dispersion, $\sigma$, around the mean of the distribution, which is given in each panel.
Since the scatter of background noise for each light profile could affect the quantity, we calculate the error for $\sigma$ by assuming the maximum deviation from the best-fit value.
The results do not change in other filter bands, as long as the rest-frame $V$- or longer wavelength bands are used.

\begin{figure*}
\figurenum{13}
\begin{center}
\epsscale{0.8}
\includegraphics[width=\textwidth,bb=0 0 288 144]{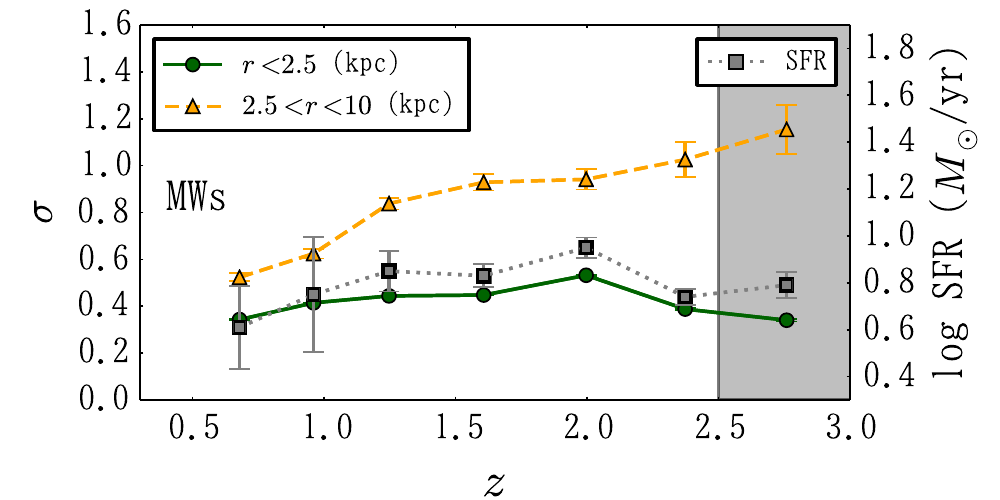}
\includegraphics[width=\textwidth,bb=0 0 288 144]{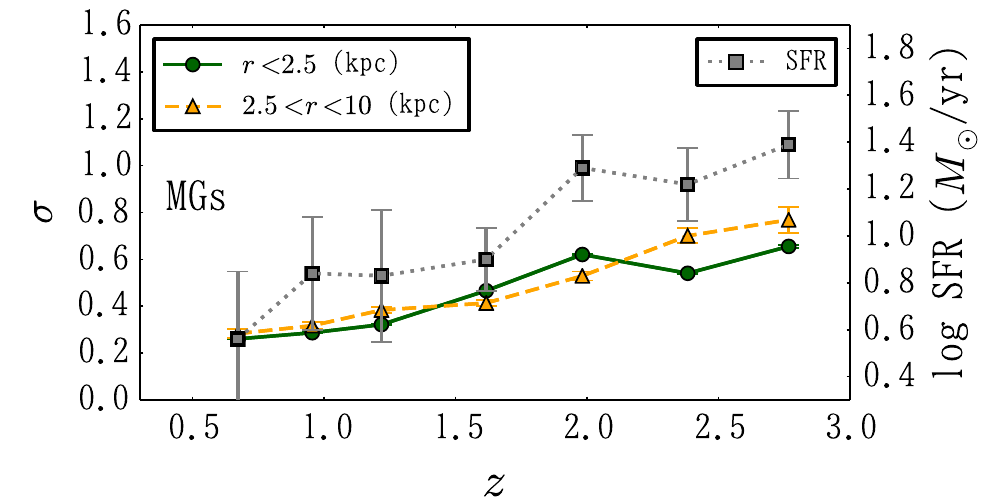}
\caption{
Evolutions of morphological variety, $\sigma$, of the $r<2.5$~kpc (green solid line with circles) and $2.5<r<10$~kpc (yellow dashed line with triangles) regions along with redshift for MWs (top) and MGs (bottom).
The error for $\sigma$, which is calculated based on the photometric error in F160W image, is negligible and smaller than the symbol size.
The sample in the shade region ($2.5<z<3.0$) of the top panel (MWs) has weaker completeness (but still $>75\%$) and might be biased toward more luminous galaxies (and higher $\sigma$), although excluding the data point does not change the trend.
Average star-formation rates (gray squares with dotted lines) are derived from the 3D-HST catalog.
The error bar for the SFR represents its dispersion.
}
\label{fig_rad4_all}
\end{center}
\end{figure*}

The evolution of $\sigma$ over the redshift range is shown in Fig.~\ref{fig_rad4_all}.
In the figure, we see that both populations have their own characteristic morphological variety.
The error bars include both of the photometric error and the scatter of background noise.
For MWs, $\sigma$ of the bulge peaks around $z\sim2$, and moderately decreases toward the lower redshift, whereas at the outer part $\sigma$ remains large ($\sim0.6$).
This difference suggests that the bulge of MWs has formed at $z>2$, while star-formation at the outer part has continued rather randomly over the entire redshift.
This might lead us to an understanding of the self-similar evolution, where the bulge mass grows by a factor of $\sim7$ from $z>2$ to $z\sim0.7$ (Section~\ref{sec:sec_42}).
To interpret the results, some secular process in the galaxies, such as stellar migration, would be needed.
MGs have, on the other hand, the peak at $z\sim2.8$ (or higher redshift), and the morphological variety rapidly calms down to $\sim0.2$ at $z\sim0.7$, which is much smaller than those for MWs.
The finding suggests a very quick transition into morphological similarity, in entirety, after the formation of the bulge at $z\sim2$.

The results above are well explained in the physical context by comparing the evolution of average SFRs with those of $\sigma$ in Fig.~\ref{fig_rad4_all}.
The median SFRs are estimated from the SED fitting for each sample in the 3D-HST catalog, and summarized in Table~\ref{tb:tb_phys}.
The SFRs for MWs and MGs peak at $z\sim2.0$ and $z\sim2.8$, respectively, which are both consistent with the result of the variety peaks.
It would be reasonably explained by the fact that the galaxy morphology begins to have a variety when the star-formation activity becomes high at, for example, clumpy regions.
Although our study is based on the stellar mass, it would not be harsh to speculate that the peak redshift also depends on the halo mass of the host galaxy; the SFR of the galaxy in massive halos peak at higher redshift and rapidly decline, whereas in smaller ones it peaks at lower redshift and slowly declines (Behroozi et al.~\citeyear{behroozi13_770}; Moster et al.~\citeyear{moster13}; McDermid et al.~\citeyear{mcdermid15}).
The evolution of the morphological variations is consistent with the results of the quenching diagnosed by the $UVJ$ colors (Sections \ref{sec:sec_43} and \ref{sec:sec_52}).
The star-formation in disk-like (younger) galaxies randomly occurs in both inner and outer regions, mainly driven by gas accretion.
The galaxy merger also increases the morphological variety.
Although distinguishing which mechanism is the main driver of the morphological variety at each epoch would be beyond the present study, we note that, according to visual inspection of each galaxy image, both mechanisms do make the galaxy morphology amorphous.
Further effort is needed to investigate the variety, in conjunction with stellar mass and its environment, for example.

After quenching, there is no such significant star-formation because there is no longer cold gas accretion any more (see discussion above), and some physical mechanisms, such as the dynamical friction of stellar components, relax the system in a similar morphology.
The quenching of galaxies happens at the corresponding redshift of the appearance of morphologically well-featured galaxies at $z\sim1$ (Dickinson~\citeyear{dickinson00}; Labbe et al.~\citeyear{labbe03}; Conselice et al.~\citeyear{conselice05}; Ravindranath et al.~\citeyear{ravindranath06}).

Lastly, we stress that the present study does not focus on the variety due to any contaminants, such as galaxies that do not evolve into MWs at $z\sim0$.
Instead, the variety here represents the diversity from ``typical" galaxies at each redshift epoch selected with the constant number density, and the time evolution of the morphological diversity of the population.

\section{CONCLUSIONS}\label{sec:sec_6}
In this study we made use of the $HST$/WFC3 and Advanced Camera for Surveys multi-band imaging data taken in CANDELS and 3D-HST survey with total survey area $\sim900$~arcmin$^2$ and $5\sigma$ source magnitude $H\sim27.0$~ABmag, which allowed us to sample galaxies with $M_*>2\times10^9~M_\odot$ at $0.5<z<3.0$.
Based on the constant cumulative number density method, we extracted the progenitors of the Milky Way (MWs) and massive galaxies (MGs) at the redshift range by following the stellar mass criteria, log$(M_*/M_\odot)=10.66-0.045z-0.13z^2$ and log$(M_*/M_\odot) =11.19-0.068z-0.04z^2$, respectively (Fig.~\ref{fig:fig_zm}).
Then, we stacked them in each redshift bin to obtain the average light and color profiles with full care for the sky subtraction and the contamination of the neighboring objects (Fig.~\ref{fig_method}).
Furthermore, we investigated the noise, systematic uncertainties (e.g., offsets of sky subtraction), and correction of axis ratio, which had not been included in the previous studies.
After reducing such noise and bias, we derived the radially resolved SEDs (radial SEDs) for the stacked profiles with sufficient S/Ns to obtain the radial profiles for stellar mass and rest-frame colors.
Based on the present results, we discussed not only the general properties and the morphological diversity for both populations, but also possible mechanisms for cessation of star-formation activities (``quenching") and stellar mass growth at later epoch.
Our conclusions are as follows;

1. MWs accumulated the stellar mass in a self-similar way in the bulge and disk. 
After its quenching, the bulge contributed to grow by $\sim5\times10^9M_\odot$ after $z\sim1$.

2. On the other hand, MGs accumulated the stellar mass in an inside-out way, obtaining more than $75\%$ of the total stellar mass at the outer part ($>2.5$~kpc) after the rapid formation phase of the massive bulge ($M_*\sim4\times10^{10}M_\odot$) at $z>2$ (Figs.~\ref{fig:fig_mass_mw}, \ref{fig:fig_mass_mw_cum} and \ref{fig_ms_inout}).

3. Quenching was found to occur strongly, depending on the stellar (and halo) masses of galaxies and bulges.
The finding suggests the evidence of bulge-related quenching mechanisms, such as morphological quenching and termination of gas flow by shock heating (Figs.~\ref{uvj9} and \ref{uv9}).

4. By comparing the median and individual light profiles, we evaluated the evolution of the varieties of galaxy morphology for the first time.
The varieties are relevant to the observed star-formation activities (SFRs and rest-frame colors) and the appearance of morphologically well-featured galaxies (Fig.~\ref{fig_rad4_all}).

In this study, we investigated the evolution of the general properties and diversity for the two populations, the MW-like and massive galaxies, in view of their progenitors as a function of the age of the universe.
The populations have the different epoch of bulge quenching, inside-out and self-similar growth of the stellar mass, similarity and variety of the morphology, and manifestation of the two main structures (bulge and disk) while undermining star-formation.
The corroboration for the origins giving rise to such dichotomy would provide us clues to further understanding of galaxy evolution in the unified framework.

\acknowledgments
We thank an anonymous referee for the valuable and constructive comments.
We would like to acknowledge Anton Koekemoer and Stijn Wuyts for their very helpful comments on data assessment and the pixel SED fitting.
This work is based on observations taken by the CANDELS Multi-Cycle Treasury Program with the NASA/ESA $HST$ and on observations taken by the 3D-HST Treasury Program (GO 12177 and 12328) with the NASA/ESA $HST$, both of which are operated by the Association of Universities for Research in Astronomy, Inc., under NASA contract NAS5-26555.
This work has been financially supported by a Grant-in-Aid for Scientific Research (24253003) of the Ministry of Education, Culture, Sports, Science and Technology in Japan. 
T.M. acknowledges support from the Japan Society for the Promotion of Science (JSPS) through JSPS research fellowships for Young Scientists.


\section*{Appendix A}
\section*{Systematic Error Caused by Sky Subtraction}
Since SED fitting derives the best-fit parameters based on spectral colors, the wrong sky subtraction on each filter image would give significantly biased results.
Although space based imaging data, compared to the ground-based one, rarely suffers from the background light, the zodiacal light still seems to have a nonnegligible effect.
To confirm whether the sky background value is appropriately estimated, we conduct the following test.
First, we estimate the sky background value at different radii, 75, 100, and 125, with a thickness of 25, in unit of pixel, for stacked 1-D profiles.
(For details about the sky subtraction for each unstacked sample, please see the main text.)
Then, we compare the estimated those sky values, including no-sky subtraction (original stacked image), and adopt the median value as the best sky value (sky$_\mathrm{best}$) and the farthest value from sky$_\mathrm{best}$ as the maximum error ($\Delta$sky).

\begin{figure}
\figurenum{A1}
\begin{center}
\includegraphics[width=8cm,bb=0 0 576 432]{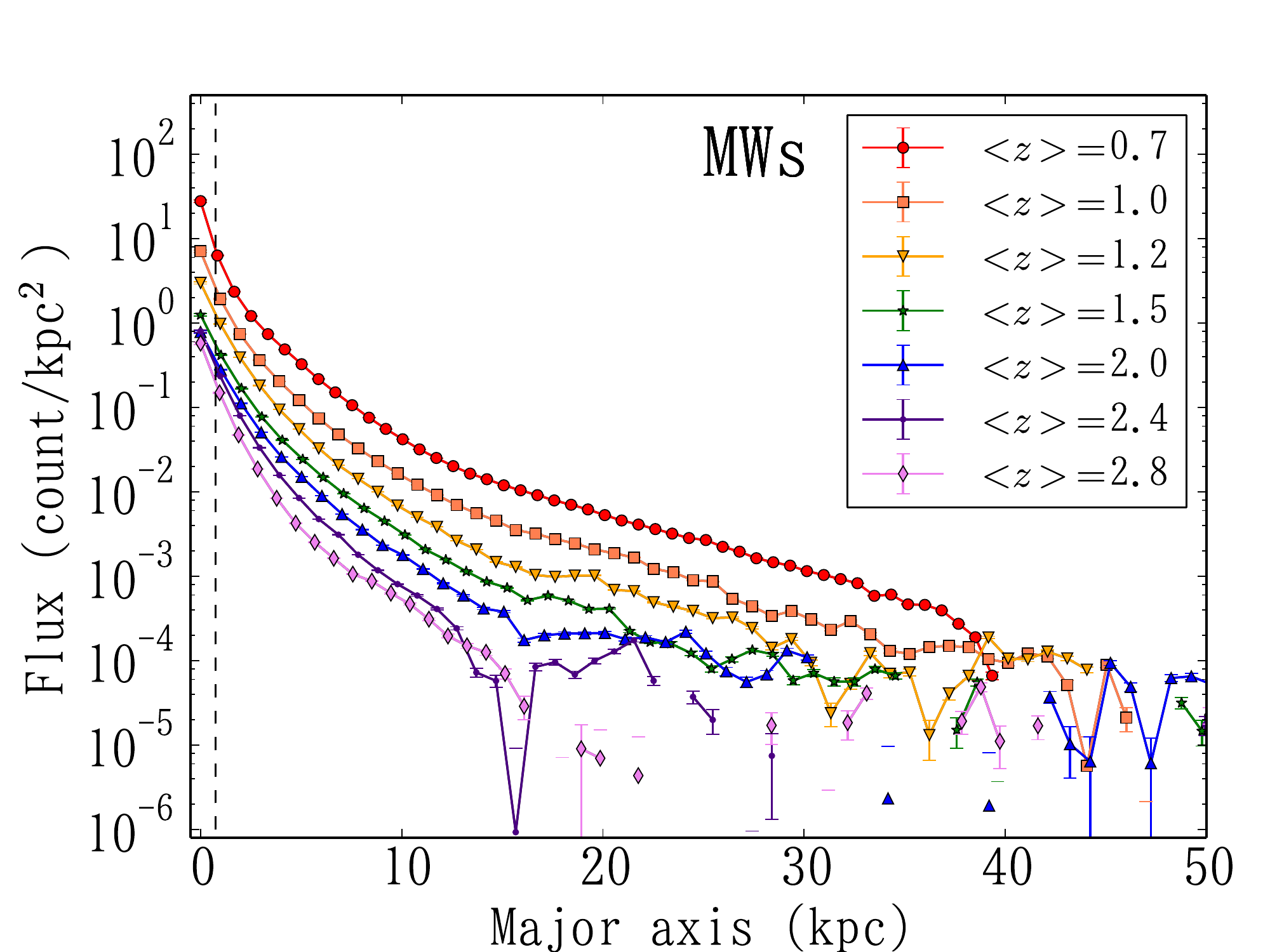}
\includegraphics[width=8cm,bb=0 0 576 432]{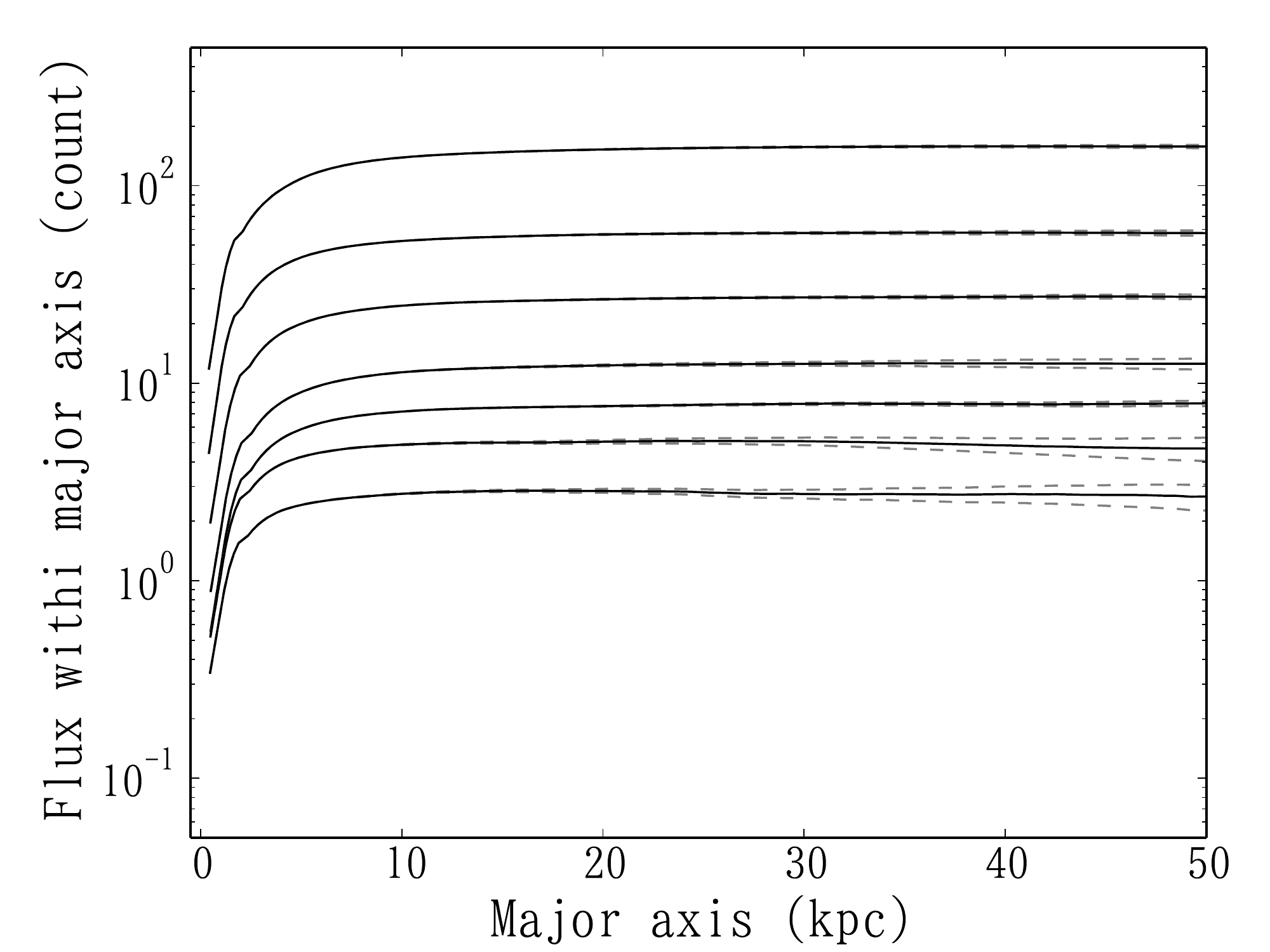}
\caption{
Top: radial light (F160W) profiles of the MWs.
The vertical axis is the observed flux, which is not corrected for the redshift effect.
The error bars are calculated by combining Eq.~(\ref{eq:eq_er}) in the main text and $\Delta$sky.
Bottom: cumulative light profiles of the same sample, for $z\sim0.7$--2.8 from top to bottom.
Profiles are subtracted with the median sky values in the caption and are shown as solid lines, while those subtracted and added the largest sky values are dashed lines.
}
\label{fig:fig_sky_mw}
\end{center}
\end{figure}

\begin{figure}
\figurenum{A2}
\begin{center}
\includegraphics[width=8cm,bb=0 0 576 432]{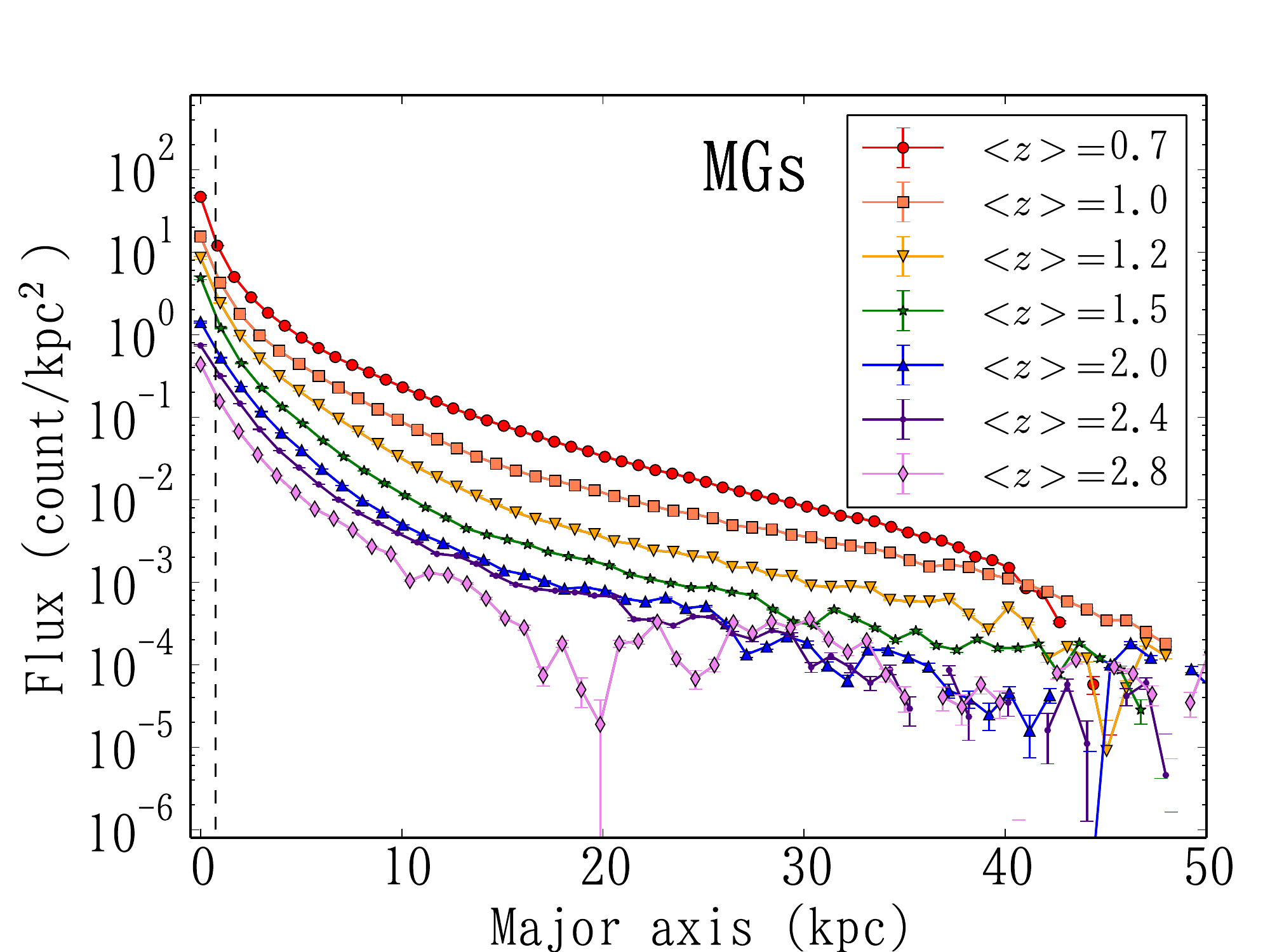}
\includegraphics[width=8cm,bb=0 0 576 432]{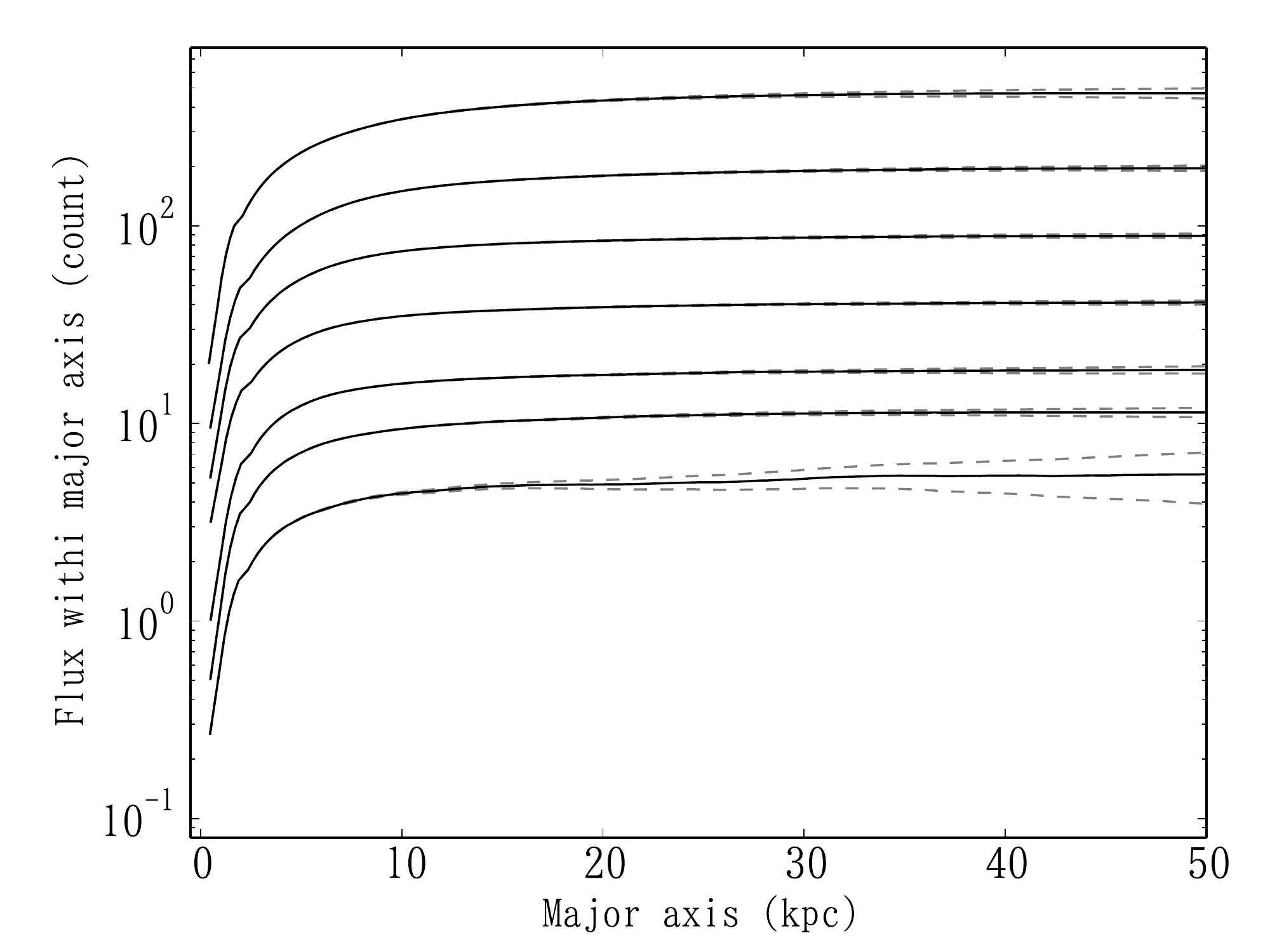}
\caption{
Same as \ref{fig:fig_sky_mw}, but for MGs.
}
\label{fig:fig_sky_me}
\end{center}
\end{figure}

We show the MWs and MGs light profiles (F160W) in Figs.~\ref{fig:fig_sky_mw} and \ref{fig:fig_sky_me}, respectively.
In the top panels, we show the radial light profiles with error bars.
The error bars include both of data error (Eq.~\ref{eq:eq_er}) and offset originated from $\Delta sky$.
The bottom panels show the cumulative light profiles of the best sky subtracted (solid) and those with shifts for $\Delta sky$ (dashed).
The differences between the radial profiles at $<20$~kpc are small, even for the samples at $z>2$, where the sky background noise would be dominant.
Since our analysis of the radial SED uses the inner parts ($<20$~kpc at $z<2$ and $<10$~kpc at $z>2$) because of the S/N limit, we conclude that our sky estimation is robust and the variation of the values does not change the final results at all.

\section*{Appendix B}
\section*{Error Estimation in Stellar Mass}
\begin{figure}
\figurenum{B1}
\begin{center}
\includegraphics[width=8cm,bb=0 0 576 432]{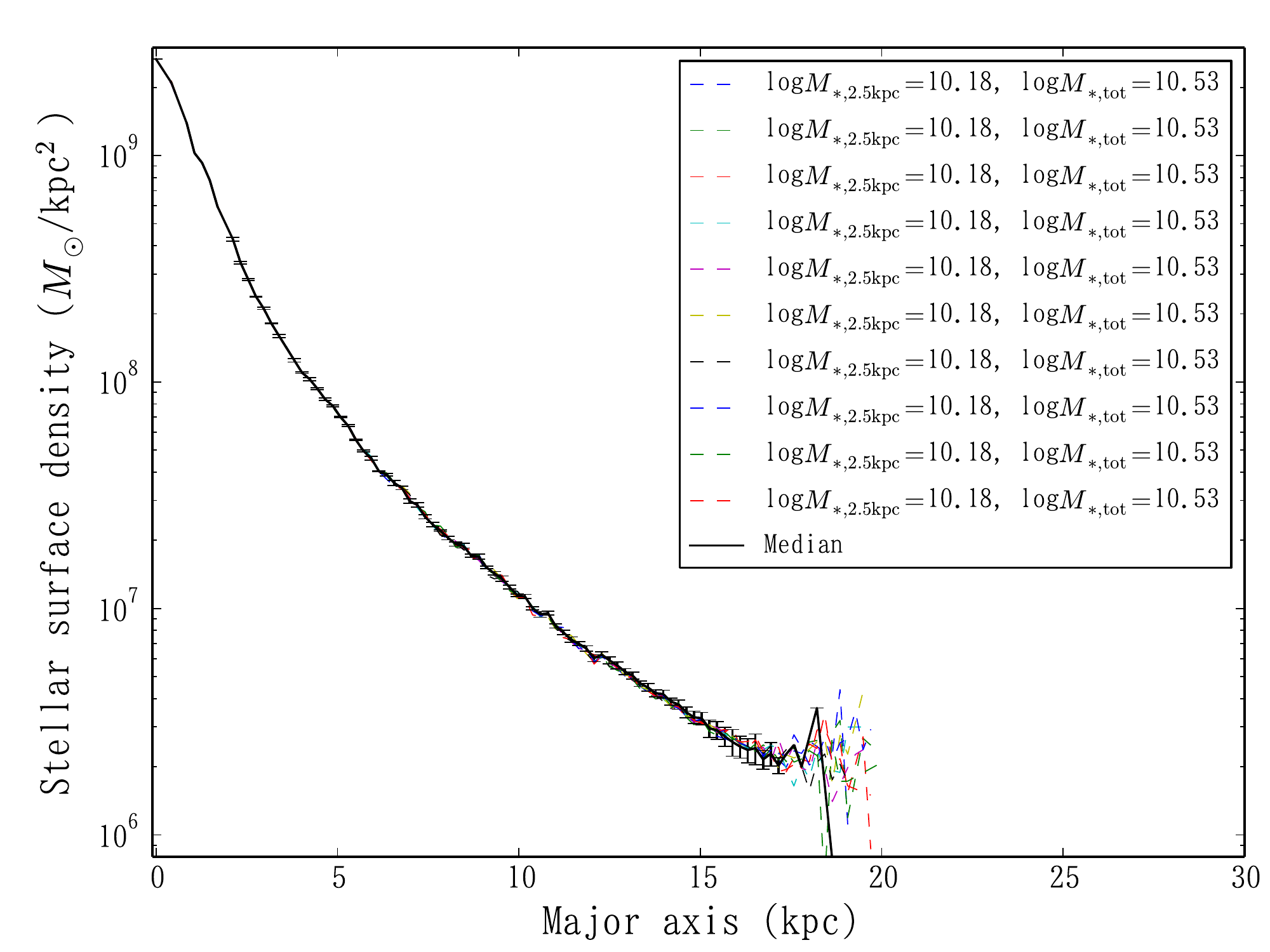}
\includegraphics[width=8cm,bb=0 0 576 432]{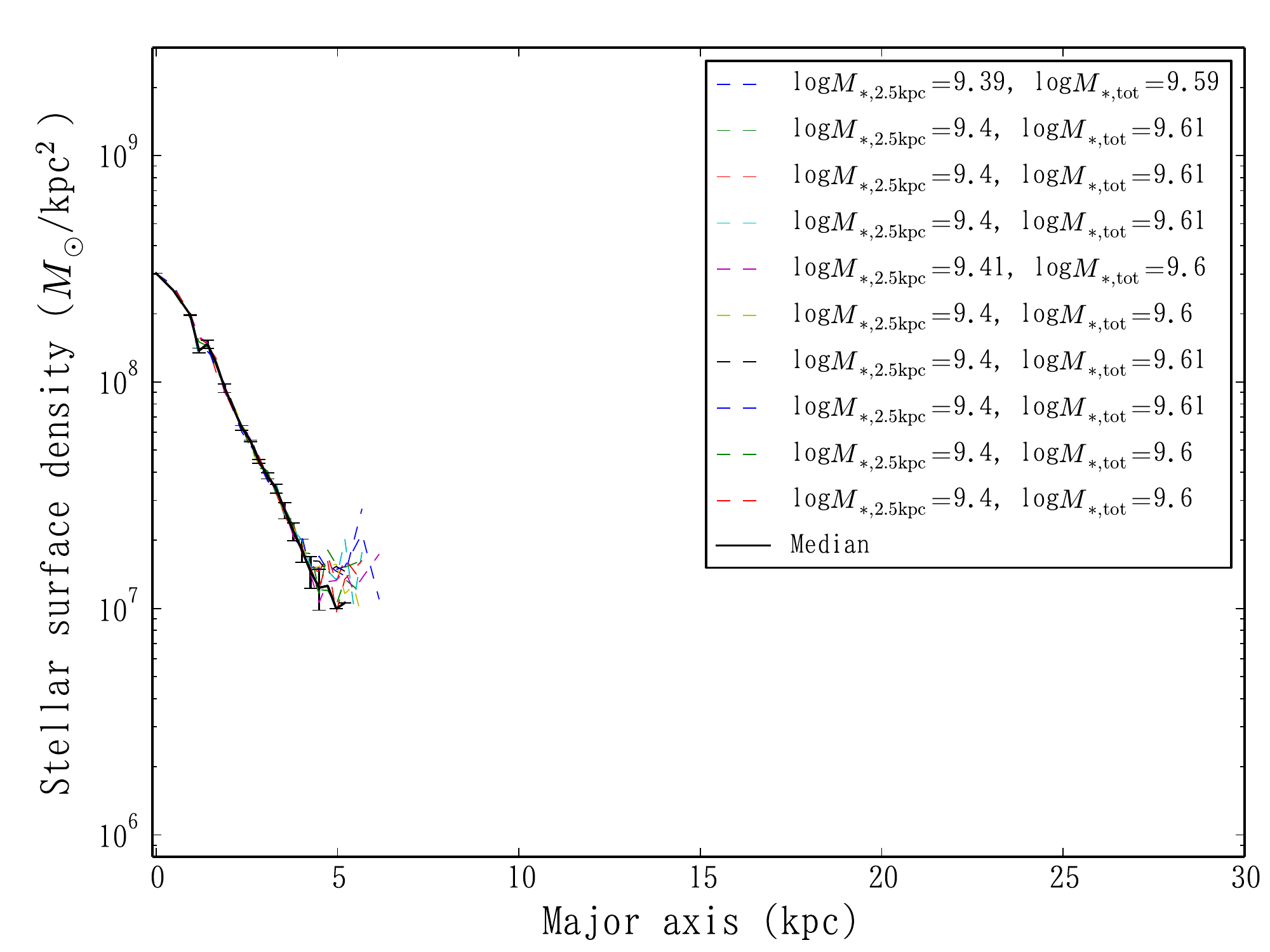}
\caption{
Examples of 10 sets of stellar mass profiles (dotted lines) derived after Monte Carlo tests for MWs at $z\sim0.7$ (top) and $z\sim2.8$ (bottom).
The solid lines with error bars are the median profile and the median absolute deviation at each radius.
The PSF effect is not corrected.
}
\label{fig:fig_mc}
\end{center}
\end{figure}

We here estimate the error in the stellar mass derived by radial SED with the criterion of S/N$>3$ for F814W, F125W and F160W.
Since the SED fit includes several parameters that would be degenerate, we estimate the errors with the Monte Carlo test.
In the test, we randomly generate the error based on the gaussian distribution with $\sigma$, which is estimated as Eq.~(\ref{eq:eq_er}), for each filter band.
Then, we shift the observed fluxes for the errors above, and derive the best-fit SED for them.
We repeat this 500 times for each pixel, and estimate the root mean square of the best-fit parameter, which we define as the error of the parameters.
Examples of 10 profiles for MWs at $\langle z\rangle=0.7$ and $\langle z\rangle=2.8$ are shown in Fig.~\ref{fig:fig_mc}, although we do not correct the PSF effect for them.
Using the criterion of S/Ns $>3$, we find that the error is negligibly small ($<3\times10^4M_\odot$ at $z\sim0.7$ and $<3\times10^5M_\odot$ at $z\sim2.8$), and hardly changes our final results.
Therefore, in the main text we adopt the scatters calculated from the different star-formation histories (exponential decline, delayed, and truncated) as the typical errors for stellar mass and rest-frame colors.

\clearpage

\begin{turnpage}
\begin{deluxetable}{ccccccccc}
\tabletypesize{\scriptsize}
\tablecolumns{9}
\tablewidth{0pc}
\tablecaption{\label{tb:tb_stack}
Number of Milky Way and Massive Galaxy Progenitors for Each Redshift Bin.}
\tablehead{
\colhead{Class} &\colhead{Band} &\colhead{$0.50\leq z <0.80$}& \colhead{$0.80\leq z <1.10$} & \colhead{$1.10\leq z <1.40$} & \colhead{$1.40\leq z <1.80$}& \colhead{$1.80\leq z <2.20$} & \colhead{$2.20\leq z <2.60$}& \colhead{$2.60\leq z <3.00$}}
\startdata

MW &F435W & 83 & 102 & 95 &97 & 85 & 192 & 131\\
 &F606W & 230 & 276 & 302 & 355 & 314 & 548 & 428\\
 &F775W & 88 & 115 & 126 & 138 & 92 & 201 & 149\\
 &F814W & 193 & 201 & 251 & 356 & 272 & 457 & 372\\
 &F850LP & 88 & 114 & 126 & 141 & 93 & 200 & 148\\
 &F125W & 261 & 313 & 357 & 496 & 374 & 597 & 470\\
 &F140W & 169 & 217 & 231 & 324 & 234 & 373 & 263\\
 &F160W & 260 & 313 & 358 & 494 & 379 & 603 & 488\\
\\
MG &F435W & 19 & 25 & 23 &20 & 30 & 29 & 10\\
 &F606W & 50 & 77 & 90 & 128 & 140 & 90 & 50\\
 &F775W & 19 & 26 & 34 & 30 & 46 & 42 & 21\\
 &F814W & 41 & 59 & 79 & 128 & 146 & 84 & 52\\
 &F850LP & 19 & 26 & 35 & 29 & 50 & 40 & 19\\
 &F125W & 56 & 81 & 102 & 166 & 221 & 147 & 83\\
 &F140W & 39 & 62 & 72 & 103 & 145 & 102 & 45\\
 &F160W & 55 & 81 & 101 & 166 & 222 & 150 & 110
\enddata
\end{deluxetable}
\end{turnpage}

\clearpage

\begin{turnpage}
\begin{deluxetable}{ccccccccc}
\tabletypesize{\scriptsize}
\tablecolumns{9}
\tablewidth{0pc}
\tablecaption{\label{tb:tb_phys}
Mean Values for the Parameters of Milky Way and Massive Galaxy Progenitors for Each Redshift Bin.
}
\tablehead{
\colhead{Class} &\colhead{Parameter} &\colhead{$0.50\leq z <0.80$}& \colhead{$0.80\leq z <1.10$} & \colhead{$1.10\leq z <1.40$} & \colhead{$1.40\leq z <1.80$}& \colhead{$1.80\leq z <2.20$} & \colhead{$2.20\leq z <2.60$}& \colhead{$2.60\leq z <3.00$}
}
\startdata

\colhead{MW} &\colhead{$\langle z \rangle$} &\colhead{0.67}& \colhead{0.96} & \colhead{1.25} &\colhead{1.60}& \colhead{2.00} & \colhead{2.39}& \colhead{2.78}\\
& \colhead{$\langle $log$M_* \rangle$~(log$M_\sun$)} &\colhead{10.57}& \colhead{10.49} & \colhead{10.40} & \colhead{10.25}& \colhead{10.04} & \colhead{9.79}& \colhead{9.51}\\
& \colhead{$\langle $log$SFRs \rangle$~(log$M_\sun$yr$^{-1}$)} &\colhead{0.61}& \colhead{0.75} & \colhead{0.85} & \colhead{0.83}& \colhead{0.95} & \colhead{0.74}& \colhead{0.79}\\
\\
\colhead{MG} &\colhead{$\langle z \rangle$} &\colhead{0.69}& \colhead{0.95} & \colhead{1.23} & \colhead{1.61}& \colhead{1.99} & \colhead{2.39}& \colhead{2.78}\\
& \colhead{$\langle $log$M_* \rangle$~(log$M_\sun$)} &\colhead{11.11}& \colhead{11.08} & \colhead{11.01} & \colhead{10.96}& \colhead{10.89} & \colhead{10.79}& \colhead{10.68}\\
& \colhead{$\langle $log$SFRs \rangle$~(log$M_\sun$yr$^{-1}$)} &\colhead{0.56}& \colhead{0.84} & \colhead{0.83} & \colhead{0.90}& \colhead{1.29} & \colhead{1.22}& \colhead{1.39}
\enddata
\end{deluxetable}
\end{turnpage}
\clearpage

\begin{deluxetable}{lcr}

\tabletypesize{\scriptsize}

\tablecaption{\label{tb:tb_sex}
SExtractor parameters for the object mask and sky estimation.
}

\tablewidth{0pt}

\tablecolumns{3}

\tablehead{
Parameter & Value & Unit
}

\startdata
DETECT\_MINAREA & 20 & pixel\\
DETECT\_THRESH & 1 &$\sigma$\\
DEBLEND\_NTHRESH & 1 &\\
DEBLEND\_MINCONT & 1 &\\
BACK\_SIZE & 250 & pixel\\
BACK\_FILTERSIZE & 1,1 & \\
BACKPHOTO\_TYPE & GLOBAL & \\
BACKPHOTO\_THICK & 5 & pixel

\enddata
\end{deluxetable}

\bibliography{adssample}

\end{document}